\documentclass[aps,showpacs]{revtex4} 

\usepackage{amssymb} 
\usepackage{graphicx} 
\usepackage{bm} 
\usepackage[T1]{fontenc} 
\usepackage{euscript} 
\usepackage{mathrsfs}

\begin{document} 

\newcommand{\cequ}{\begin{eqnarray}} 
\newcommand{\fequ}{\end{eqnarray}} 
\newcommand{\anticomut}[2]{\left\{#1,#2\right\}} 
\newcommand{\comut}[2]{\left[#1,#2\right]} 
\newcommand{\comutd}[2]{\left[#1,#2\right]^{*}} 

\title{\bf 
Hamiltonian thermodynamics of $d$-dimensional ($d\geq4$) 
Reissner-Nordstr\"om anti-de Sitter black holes 
with spherical, planar, and hyperbolic topology} 
\author{Gon\c{c}alo A. S. Dias\footnote{Email: gadias@fisica.ist.utl.pt}} 
\affiliation{Centro Multidisciplinar de Astrof\'{\i}sica - CENTRA \\ 
Departamento de F\'{\i}sica, Instituto Superior T\'ecnico - IST,\\ 
Universidade T\'ecnica de Lisboa - UTL,\\ 
Avenida Rovisco Pais 1, 1049-001 Lisboa, Portugal} 
\author{Jos\'e P. S. Lemos\footnote{Email: lemos@fisica.ist.utl.pt}} 
\affiliation{Centro Multidisciplinar de {}Astrof\'{\i}sica - CENTRA \\ 
Departamento de F\'{\i}sica, Instituto Superior T\'ecnico - IST,\\ 
Universidade T\'ecnica de Lisboa - UTL,\\ 
Avenida Rovisco Pais 1, 1049-001 Lisboa, Portugal} 
\begin{abstract} 
The Hamiltonian thermodynamics formalism is applied to the general
$d$-dimensional Reissner-Nordstr\"om-anti-de Sitter black hole with
spherical, planar, and hyperbolic horizon topology. After writing its
action and performing a Legendre transformation, surface terms are
added in order to guarantee a well defined variational principle with
which to obtain sensible equations of motion, and also to allow later
on the thermodynamical analysis. Then a Kucha\v{r} canonical
transformation is done, which changes from the metric canonical
coordinates to the physical parameters coordinates. Again a well
defined variational principle is guaranteed through boundary
terms. These terms influence the fall-off conditions of the variables
and at the same time the form of the new Lagrange
multipliers. Reduction to the true degrees of freedom is performed,
which are the conserved mass and charge of the black hole. Upon
quantization a Lorentzian partition function $Z$ is written for the
grand canonical ensemble, where the temperature $\bf T$ and the
electric potential $\phi$ are fixed at infinity. After imposing
Euclidean boundary conditions on the partition function, the
respective effective action $I_*$, and thus the thermodynamical
partition function, is determined for any dimension $d$ and topology
$k$. This is a quite general action.  Several previous results can be
then condensed in our single general formula for the effective action
$I_*$. Phase transitions are studied for the spherical case, and it is
shown that all the other topologies have no phase transitions. A
parallel with the Bose-Einstein condensation can be
established. Finally, the expected values of energy, charge, and
entropy are determined for the black hole solution.
\end{abstract} 


\maketitle 

\section{Introduction} 
\label{introduction} 

There are many approaches to calculate the entropy and the 
thermodynamics of a black hole. One can follow the original route 
where methods of field second quantization in a collapsing object are 
used to calculate the temperature $\bf T$ \cite{hawking1}, and then 
uses the black hole laws to find the corresponding entropy 
\cite{bardeencarterhawking1973}.  Or one can use the Euclidean path 
integral approach to quantum gravity \cite{hartlehawking,hawking2} and 
its developments \cite{york1}, to obtain those thermodynamic properties 
\cite{york1,braden,zaslavskii1,hawkingpage,peca1,peca2,witten,cejm1,cejm2} 
(see also \cite{grossperryyaffe} and compare with \cite{york1}). 
There are still other methods.  The method we follow here is the one 
that builds a Lorentzian Hamiltonian classical theory of the gravity 
field and possibly other fields, and then obtain a Lorentzian time 
evolution operator in the Schr\"odinger picture. Afterward one 
performs a Wick rotation from real to imaginary time, in order to find 
a well defined partition function.  The prescription implicit in this 
approach is, first, find the Hamiltonian of the system, second, 
calculate the time evolution between a final state and an initial 
state, i.e., between the final bra vector state $<g_2,\Sigma_2,t_2|$, 
with metric $g_2$ on a spatial hypersurface $\Sigma_2$ at some generic 
prescribed time $t_2$, and the initial ket vector state, 
$|g_1,\Sigma_1,t_1>$, with metric $g_1$ on a spatial hypersurface 
$\Sigma_1$ at some generic prescribed time $t_1$, and, third, 
Euclideanize time. Here, the amplitude to propagate to a configuration 
$<g_2,\Sigma_2,t_2|$ from a configuration $|g_1,\Sigma_1,t_1>$ , is 
represented by $<g_2,\Sigma_2,t_2| 
\exp\left(-iH(t_2-t_1)\right)|g_1,\Sigma_1,t_1>$ in the Schr\"odinger 
picture.  Euclideanizing time, $t_2-t_1=-i\beta$ and summing over a 
complete orthonormal basis of configurations $g_n$ one obtains in 
general the partition function $Z=\sum \exp\left[-\beta\left( 
E_n-q_n\phi\right)\right]$, of the field $g$ at a temperature $\textbf{T} 
\equiv1/\beta$ and at some chemical potential $\phi$, where $E_n$ is 
the eigenenergy of the eigenstate $g_n$, and $q_n$ is the eigenvalue 
of some variable conjugate to $\phi$, such as the particle number 
(when $\phi$ is effectively the chemical potential), or the charge 
number (when $\phi$ is the electric potential), for instance.  This 
route is based on the Hamiltonian methods for covariant gravity 
theories \cite{dirac,adm,rt} and for black hole spacetimes on the 
Hamiltonian method of Kucha\v{r} \cite{kuchar}. It was developed by 
Louko and Whiting in \cite{louko1} for the specific problem of finding 
black hole entropies and thermodynamic properties. 

Indeed Kuchar \cite{kuchar} applied his procedure to the full vacuum 
Schwarzschild black hole spacetime, which is really composed of a 
spherically symmetric white hole plus a black hole plus two 
asymptotically flat regions. These regions are well pictured in a 
Carter-Penrose diagram.  By considering the spacelike foliations of 
the full manifold, the true dynamical degree of freedom of the phase 
space of the Schwarzschild black hole was found. This degree of 
freedom, represented by one pair of canonical variables, is composed 
of the mass $M$ of the solution and its conjugate momentum, which 
physically represents the difference between the Killing times at 
right and left spatial infinities. Louko and Whiting \cite{louko1}, by 
adapting the formalism to a spacelike foliation to the right of the 
future event horizon of the solution, and imposing appropriate 
boundary conditions, obtained the Hamiltonian $H$, the time evolution 
operator $\exp(-iHt)$ in the Schr\"odinger picture, the associated 
partition function and finally the thermodynamics. This method has 
been applied for various theories of gravity, in several different 
dimensions, and in either asymptotically flat or asymptotically AdS 
spacetimes. 

There are several applications of the formalism to four-dimensional
spacetimes \cite{louko1,louko3,louko4}, Now, it is clear that it is
important to understand the physics in several different dimensions.
Indeed, hints from many places like string theory, AdS/CFT (Anti de
Sitter/conformal field theory) conjecture, extra large dimensions and
the connected braneworld scenarios, point out to the possibility of a
world with other space dimensions.  A first attempt to study the
thermodynamics of a higher dimensional black hole using the
Hamiltonian approach was done for $d=5$ Lovelock gravity
\cite{louko5}. A related formalism developed in
\cite{kunst1,kunst2} studied some higher dimensional black
holes that admit a reduction to a two-dimensional dilaton-gravity
theory.  There are also incursions in the application of the formalism
into lower dimensional dilaton-gravity theories, e.g., in $d=2$
\cite{louko2}, and in $d=3$ \cite{dl3,dl4}.  In \cite{bose} a
connection between the Hamiltonian formalism of \cite{louko1} and the
path-integral formalism of \cite{york1} is made.

So here, we are interested in applying the Hamiltonian formalism 
introduced by Louko and Whiting \cite{louko1}, (see also 
\cite{louko3,louko4,louko2,kunst1,kunst2,dl3,dl4,louko5,bose}), 
to the $d$-dimensional, $d\geq4$, Reissner-Nordstr\"om black holes 
with spherical, planar, and hyperbolic horizon compact topology, and 
with a negative cosmological constant, i.e., in an asymptotically 
anti-de Sitter (AdS) spacetime.  Spherically symmetric electrically 
charged black holes in $d$ dimensions are also known as Tangherlini 
black holes \cite{tangherlini}, or simply $d$-dimensional 
Reissner-Nordstr\"om AdS black holes, see also \cite{mp} for the Kerr 
the $d$-dimensional Kerr black holes. Planar toroidal compact black 
holes in four dimensions were discussed in \cite{lemos1,lemos2} and 
charged ones in \cite{lemos_zanchin} (see also \cite{cai1,vanzo}). 
Hyperbolic toroidal compact black holes in four dimensions were 
studied in \cite{mann}, and the charged version of the three together 
in \cite{louko4}. In $d$ dimensions these topological black holes were 
analyzed in \cite{birmingham}, see also 
\cite{cai2,santos_dias_lemos}. The solution for Kerr-AdS black holes 
in $d$ dimensions was found in \cite{kerrantidesitter}. One notes that 
if the cosmological constant is nonnegative, black holes with 
nonspherical topology do not exist. 

To develop the Hamiltonian formalism for these $d$-dimensional AdS 
black holes an important issue one has to deal with is to find the 
asymptotic fall-off conditions at infinity. In four dimensions the 
problem was settled by Henneaux and Teitelboim \cite{ht}, where 
through the precise asymptotic structure of the Kerr-AdS metric and 
acting on that structure with the four-dimensional AdS group one was 
able to find the correct fall-off conditions.  Louko and Winters-Hilt 
\cite{louko3} used these fall-off conditions to study the 
four-dimensional Reissner-Nordstr\"om-AdS black hole. In $d$ 
dimensions no such procedure is available, although the Kerr-AdS 
metric solution in $d$ dimensions has already been found 
\cite{kerrantidesitter}.  Thus, by following the same procedure as in 
\cite{ht} one should be able to arrive at the correct fall-off 
conditions. Fortunately, in our case the problem is much simpler and 
we can adapt with some care to $d$ dimensions the fall-off conditions 
of \cite{louko3}.  Then, after following other procedures, one finds 
the reduced Hamiltonian, and a statistical analysis can finally be 
performed. We study the grand canonical ensemble and find a very 
general effective action $I_*$, or equivalently, a very general 
partition function.  From the action $I_*$ one can 
readily extract all the relevant thermodynamical information.  We 
study the phase transitions of the system and the thermodynamic 
quantities, such as temperature and entropy, of the solution 
containing a black hole. The phase transitions are between hot flat 
space and black holes, two different sectors of the solution space, 
one with trivial topology, the other with black hole topology, 
respectively.  These phase transitions show similarities with the 
Bose-Einstein condensation phenomenon. 

There are several particular results that can be obtained from our
general action formula $I_*$. First, within works using a Hamiltonian
thermodynamics formalism we recover the thermodynamics for spherically
symmetric Reissner-Nordstr\"om-AdS black holes in four dimensions
($d=4$) \cite{louko3}, the thermodynamics for AdS black holes with
planar and hyperbolic topology in four dimensions ($d=4$)
\cite{louko4}, and the results for the spherical black hole in a
finite box with radius $r_{\rm B}$ (when $r_{\rm B}\rightarrow\infty$)
in a Lovelock $d=5$ theory, which can also be called a Gauss-Bonnet
theory, without a cosmological constant \cite{louko5}.  Second, within
works using an Euclidean path integral method we also recover from our
general action formula, when appropriate, the thermodynamics of the
$d$-dimensional spherical Reissner-Nordstr\"om black hole studied in
\cite{cejm1,cejm2}, as well as the thermodynamics of the
$d$-dimensional Schwarzschild black hole studied in \cite{witten}.
Moreover, we obtain the thermodynamics of the Reissner-Nordstr\"om-AdS
black holes in four dimensions ($d=4$) with spherically symmetric
studied in \cite{peca1}, which in turn recovers results found in
\cite{york1,braden,zaslavskii1} and in \cite{hawkingpage}, as well as
the thermodynamics obtained in \cite{peca2} for the charged AdS black
holes in planar topology.

The paper layout is thus the following: in Sec. \ref{bh} we write the 
ansatz for the black hole solutions of the charged black holes in $d$ 
dimensions with spherical, planar toroidal, and hyperbolic toroidal 
topology, plus the ansatz for the vector potential $A$, following it 
with the metric and vector potential solutions for the $d$-dimensional 
black holes.  The Carter-Penrose diagram of the charged black holes in 
$d$ dimensions with spherical, planar toroidal, and hyperbolic 
toroidal topology, is then depicted. In Sec. \ref{formalism}, the ADM 
form for the metric and $A$ are spelled out, the canonical description 
for the Einstein-Maxwell action in $d$ spacetime dimensions is done 
and is followed by the reconstruction of the action.  After the 
canonical transformations, we reduce the Hamiltonian to the physical 
degrees of freedom.  There follows the canonical quantization through 
the Schr\"odinger representation of the time evolution operator. In 
Sec. \ref{thermo} the thermodynamics is studied.  First, there is the 
construction of the partition function for the grand canonical 
ensemble through the imposition of Euclidean boundary conditions on 
the time evolution operator found previously.  Then, the critical 
points of the effective action in the partition function integral are 
obtained.  Next, the effective action is evaluated at the critical 
points, where, depending on the value of the effective action at these 
critical points, one can obtain hot flat space or a black hole through 
a phase transition.  The results are compared to previous results.  It 
is then possible to determine the expected values of the energy and 
charge, plus the entropy, from the saddle point approximation of the 
partition function of the ensemble.  In Sec. \ref{conclusions} we 
conclude.  We choose $G=1$, $c=1$, $\hbar=1$, and $k_{\rm B}=1$ 
throughout. 

\section{Charged black holes in $d$ dimensions with 
spherical, planar, and hyperbolic topology in asymptotically 
AdS spacetimes} 
\label{bh} 

\subsection{Solutions with spherical, planar, and hyperbolic compact 
topology} 
\label{bh_solutions} 
The electrically charged AdS black hole solution 
can be generically written as (see \cite{santos_dias_lemos}) 
\begin{eqnarray} 
ds^2 = - F(R)\, dT^2 + F(R)^{-1}\,dR^2+R^2 
(d\,\Omega_{d-2}^k)^2, 
\label{general_metric} 
\end{eqnarray} 
where we have $d\geq4$, $T$ is the Schwarzschild time coordinate, 
$R$ is the Schwarzschild time coordinate, and $F(R)$ 
is a function that yields 
horizons and thus black holes. Its form depends on the 
theory, we will be interested in $d$-dimensional general 
relativity. 
The constant $k$ has values $k=1,0,-1$ whether the topology 
is spherical, planar toroidal, or hyperbolic toroidal, 
all three being compact topologies. 
The angular part of the metric (\ref{general_metric}) is, 
for each $k=1,0,-1$, 
\begin{eqnarray} 
\label{angular} (d\Omega_{d-2}^1)^2 \!\!\!&=& \!\!\! 
d\theta_1^2+\sin^2\theta_1\,d\theta_2^2+ \cdots 
+\prod_{i=1}^{d-3}\sin^2\theta_i\,d\theta_{d-2}^2\,, \nonumber \\ 
(d\Omega_{d-2}^0)^2 \!\!\!&=& \!\!\! d\theta_1^2+d\theta_2^2+ 
d\theta_3^2+\cdots +d\theta_{d-2}^2\,, \nonumber \\ 
(d\Omega_{d-2}^{-1})^2 \!\!\!&=& \!\!\! 
d\theta_1^2+\sinh^2\theta_1\,d\theta_2^2\!+\cdots+\! 
\sinh^{2}\theta_1\!\!\prod_{i=2}^{d-3}\sin^2\theta_i\, 
d\theta_{d-2}^2\,. 
\end{eqnarray} 
In the spherical case, $k=1$, one can take the range of coordinates as 
the usual one $0\leq\theta_1<\pi$, $0\leq\theta_i<2\,\pi$ for 
$i\geq2$.  In the planar case, $k=0$, the range is arbitrary, though 
finite for a compact surface, and we can choose 
$0\leq\theta_i<2\,\pi$.  In the hyperbolic-compact surface case, 
$k=-1$, the situation is in general more involved, and we need to 
choose a finite area surface for the black hole horizon in order to be 
able to study its thermodynamics. For example, in $d=4$ spacetime 
dimensions, the horizon is a 2-dimensional hyperbolic torus, i.e, a 
Riemann surface with genus $g_{\rm enus}\geq2$. For more details, see 
\cite{lemos1} and \cite{birmingham} (see also \cite{vanzo}). 
The vector potential one-form which is generically written 
as 
\begin{eqnarray} 
    A = A(R)\, dT\,, 
    \label{vector_potentialform} 
\end{eqnarray} 
where the electric potential $A(R)$ depends on the theory 
one is studying. We will work with general relativity in $d$-dimensions. 

\subsection{Black hole solutions: metric and vector potential} 
\label{metricandvectorpotential} 


\begin{figure} 
[b] 
\centerline{\includegraphics{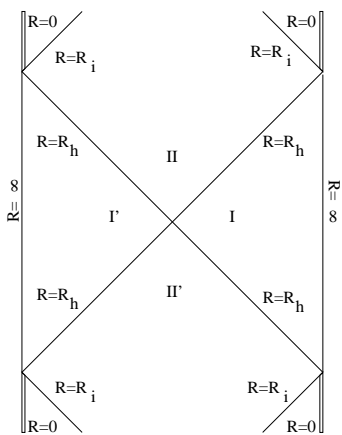}} 
\caption {{\small The Carter-Penrose diagram for the 
Reissner-Nordstr\"om-AdS black hole with 
spherical, planar toroidal, and hyperbolic toroidal topologies, where 
$R_{\textrm{h}}$ is the outer horizon radius, $R_{\rm i}$ is the inner 
horizon radius, and the double line is the timelike singularity at 
$R=0$. The static region I, from the bifurcation $(d-2)$-manifold to 
spacelike infinity $R=\infty$, is the relevant region for our 
analysis. The regions II and II' are inside the outer horizon, beyond 
the foliation domain. Region I' is the symmetric of region I, toward 
left infinity.  Note that any point in the diagram is a 
$(d-2)$-dimensional manifold. The bifurcation manifold can be a 
sphere, a planar torus, or a hyperboloidal torus, always 
$(d-2)$-dimensional.}} 
\label{cp_diagram} 
\end{figure} 

We are interested in general relativity coupled to a Maxwell 
field in $d$-dimensions 
whose action is 
\begin{equation} 
S = \int\,d^{d}x 
\sqrt{-g}\, 
\left(\frac{1}{16\pi}\,[R+(d-1)(d-2)\,l^{-2}] 
-\frac14 F_{\mu\nu}F^{\mu\nu}\right)\,, 
\label{einsteinmaxwell1} 
\end{equation} 
with 
$g$ being the determinant of the metric $g_{\mu\nu}$, 
$R$ is the Ricci scalar constructed from the 
Riemann tensor $R_{\alpha\beta\gamma\delta}$, 
$l$ is the AdS length, which the makes up 
the cosmological constant through $-3\,l^{-2}$, 
and $F_{\mu\nu}=\partial_\mu A_\nu-\partial_\nu A_\mu$ 
is the Maxwell tensor, with $A=A_\mu dx^\mu$ the 
one-form vector potential. 

The electrically charged AdS black holes 
solutions of the 
Einstein-Maxwell-AdS action in $d$ dimensions, 
have 
$F(R)$ given by 
\begin{eqnarray} 
F(R) = 
k+l^{-2}\,R^2-\frac{2 {\bar M}}{R^{d-3}}+ 
\frac{{\bar Q}^2}{R^{2(d-3)}}\:, 
\label{general_f} 
\end{eqnarray} 
Here, the quantities $\bar M$ and $\bar Q$ are the mass and 
the charge parameters, respectively.  The definition of mass in 
general relativity is always ambiguous.  Usually one defines the 
Arnowitt-Deser-Misner (ADM) mass $m$, which in principle, for 
asymptotically well defined spacetimes is a well defined quantity. The 
black holes we study here are asymptotically AdS and so have a well 
defined mass. However, depending on the values one gives for the 
cyclic coordinates of the torus, one gets different answers in the 
planar and hyperbolic toroidal cases, not so in the spherical case 
which has well defined angular coordinates. Moreover, when performing 
our canonical analysis, as it will be done here, a mass which is the 
one necessary to make the important canonical transformations, and 
renders the expressions most simple, pops up. We call this mass the 
canonical mass $M$. The relation between the canonical mass and the 
mass parameter is, as we will see, 
$M=\frac{(d-2)\,\Sigma^k_{d-2}}{8\pi}\, {\bar M}$, 
where 
$\Sigma^k_{d-2}$ is the area of the $(d-2)$ unit surface.  For 
instance, in the spherical case and for zero cosmological constant, 
one has ${\bar M}=\frac{8\pi \, m}{(d-2)\,\Sigma^{\,\,1}_{d-2}}$, 
where now in the spherical case $M$ is also the ADM mass, $M=m$, and 
$\Sigma^{\,\,1}_{d-2}$ is the area of the $(d-2)$ unit sphere 
\cite{mp}. The same applies for the electrical charge, although here 
the ambiguities are not as strong.  Here the canonical charge $Q$, 
i.e., the one necessary to make the important canonical 
transformations, and the charge parameter are equal, 
$Q={\bar Q}$. 
None of these is the ADM (Gauss) charge $q$.  For instance, in the 
spherical case and for zero cosmological constant, one has ${\bar 
Q}^2= q^2 \left( \frac{8\pi }{(d-2)\,(d-3)} \right)$. 
The vector potential function $A(r)$ is 
\begin{eqnarray} 
    A(r) &=& \sqrt{\frac{d-2}{8\pi(d-3)}}\frac{\bar Q}{R^{d-3}}\,, 
    \label{vector_potential} 
\end{eqnarray} 
where the relevant quantities have been defined above. 

The corresponding Carter-Penrose diagram is given in Fig.\ 
\ref{cp_diagram}. 

\section{Hamiltonian thermodynamics formalism} 
\label{formalism} 

\subsection{ADM form of the metric and vector potential} 
\label{adm_metric} 
The ansatz for the metric field with which we start 
our canonical analysis is given by 
\begin{equation} 
ds^2 = -N(t,r)^2 dt^2+\Lambda(t,r)^2(dr+N^r(t,r) 
dt)^2+R(t,r)^2(d\,\Omega_{d-2}^k)^2\,, 
\label{adm_ansatz} 
\end{equation} 
where $t$ and $r$ are the time and radial ADM coordinates used in the 
ADM metric ansatz for the black hole solutions of the 
Einstein-Maxwell-AdS action.  In the subsequent developments we follow 
the basic formalism developed by Kucha\v{r} \cite{kuchar}.  The 
canonical coordinates are $R$ and $\Lambda$, which are functions of 
$t$ and $r$, $R=R(t,r),\,\Lambda=\Lambda(t,r)$. Now, $r=0$ is 
generically on the horizon as analyzed in \cite{kuchar}, but for our 
purposes $r=0$ represents the horizon bifurcation $(d-2)$-manifold of 
the Carter-Penrose diagram (see, 
\cite{louko1,louko3,louko4,kunst1,kunst2,bose,louko2,dl3,dl4,louko5}). 
The coordinate $r$ tends 
to $\infty$ as the coordinates themselves tend to infinity, and $t$ is 
another time coordinate. The remaining functions are the lapse 
$N=N(t,r)$ and shift functions $N^r=N^r(t,r)$ and will play the role 
of Lagrange multipliers of the Hamiltonian of the theory.  The 
canonical coordinates $R=R(t,r),\,\Lambda=\Lambda(t,r)$ and the lapse 
function $N=N(t,r)$ are taken to be positive.  The angular coordinates 
are left untouched, due to the symmetries assumed. The ansatz 
(\ref{adm_ansatz}) is written in order to perform the foliation of 
spacetime into spacelike hypersurfaces, and thus separates the spatial 
part of the spacetime from the temporal part.  Indeed, the canonical 
analysis requires the explicit separation of the time coordinate from 
the other space coordinates, and so in all expressions time is treated 
separately from the other coordinates. It breaks explicit covariance 
of the Einstein-Maxwell-AdS theory, but it is necessary in order to 
perform the Hamiltonian analysis.  The metric coefficients of the 
induced metric on the hypersurfaces become the canonical variables, 
and the momenta are determined in the usual way, by replacing the time 
derivatives of the canonical variables, the velocities.  Then, using 
the Hamiltonian one builds a time evolution operator to construct an 
appropriate thermodynamic ensemble for the geometries of a quantum 
theory of gravity. 

The canonical description of the vector potential one-form is given by 
\begin{eqnarray} 
    A &=& \Gamma dr + \Phi dt\,, 
    \label{a_ansatz} 
\end{eqnarray} 
where $\Gamma$ and $\Phi$ are functions of $t$ and $r$, i. e., 
$\Gamma(t,r)$ and $\Phi(t,r)$.  The function $\Gamma(t,r)$ is the 
canonical coordinate associated with the electric field, and the 
function $\Phi(t,r)$ is the Lagrange multiplier associated with the 
electromagnetic constraint, which is Gauss' Law. 

\subsection{Canonical formalism} 
\label{can_form} 
We now replace the ansatz for the metric (\ref{adm_ansatz}) and 
the ansatz for the one-form vector potential 
(\ref{a_ansatz}) into the action in 
Eq.\  (\ref{einsteinmaxwell1}), obtaining 
\begin{eqnarray} 
    S_\Sigma[\Lambda,\,R,\,\Gamma;N,\,N^r,\,\Phi] &=& 
    \int\,dt\,\int_0^\infty\,dr\;a\,\left\{k\,B\,N\Lambda R^{d-4}+ 
    6\,l^{-2}N\Lambda R^{d-2}-2(d-2)N^{-1}\dot{R}\dot{\Lambda} 
    R^{d-3}\right. \nonumber \\ 
    && -BN^{-1}\Lambda\dot{R}^2R^{d-4} 
    + 2(d-2)N^{-1}\dot{R}\left(\Lambda N^r\right)'+2(d-2)N^{-1} 
    \dot{\Lambda}\left(R' N^r\right)\nonumber \\ 
    && -2(d-2)N^{-1}\left(\Lambda N^r\right)' 
    \left(R' N^r\right)R^{d-3}+ 
    2B N^{-1}R'\dot{R}N^r\Lambda R^{d-4}-B N^{-1}\Lambda 
    \left(R'N^r\right)^2R^{d-4} \nonumber \\ 
    && -2(d-2)N\left(\Lambda^{-1}\right)'R'R^{d-3} 
    -B N\Lambda^{-1}\left(R'\right)^2R^{d-4}-2(d-2)N\Lambda^{-1}R'' 
    R^{d-3}\nonumber \\ 
    && \left. +8\pi N^{-1}\Lambda^{-1} 
    \left(\dot{\Gamma}-\Phi'\right)^2R^{d-2}\right\}\,, 
\end{eqnarray} 
where $B$ is defined as 
$B=(d-3)(d-2)$, $a$ is defined as 
\begin{equation} 
a=\frac{\Sigma^k_{d-2}}{16\pi}\,, 
\label{areal} 
\end{equation} 
and, e. g., for $k=1$ we have 
$\Sigma^1_{d-2}=2\pi^{(d-1)/2}/\Gamma(\frac{d-1}{2})$, 
$\Gamma(x)$ being the Gamma function. 

The conjugate momenta are 
\begin{eqnarray} 
    P_\Lambda &=& -2\,a\,(d-2)R^{d-3}N^{-1} 
    \left(\dot{R}-N^rR'\right)\,, 
    \label{p_lambda}\\ 
    P_R &=& -2\,a\,(d-2)R^{d-4}N^{-1} 
    \left[R\left(\dot{\Lambda}-\left(\Lambda N^r\right)' 
    \right)+(d-3)\Lambda\left(\dot{R}-N^rR'\right)\right]\,, 
    \label{p_r}\\ 
    P_\Gamma &=& 16\pi\,a\,N^{-1}\Lambda^{-1} 
    \left(\dot{\Gamma}-\Phi'\right)R^{d-2}\,. 
    \label{p_gamma} 
\end{eqnarray} 
After a Legendre transformation we find 
\begin{eqnarray} 
    S_\Sigma\left[\Lambda,\,R,\,\Gamma,\,P_\Lambda,\,P_R,\,P_\Gamma; 
    N,\,N^r,\,\tilde{\Phi}\right] &=& \int\,dt\,\int_0^\infty\,dr 
    \left(P_\Lambda\dot{\Lambda}+P_R\dot{R}+P_\Gamma\dot{\Gamma} 
    -NH-N^rH_r-\tilde{\Phi}G\right)\,, 
    \label{ham_action} 
\end{eqnarray} 
where $\tilde{\Phi}$ is defined as 
$\tilde{\Phi}\equiv\Phi-N^r\Gamma$, and 
the constraints are 
\begin{eqnarray} 
    H &=& \frac{(d-3)}{4\,a(d-2)}\Lambda P_\Lambda^2R^{-(d-2)}+ 
    \frac{1}{16\pi\,a}\Lambda P_\Gamma^2 R^{-(d-2)}- 
    \frac{1}{2\,a(d-2)}P_RP_\Lambda R^{-(d-3)}\nonumber \\ 
    && +\,a\,\left(-k\,B\,\Lambda R^{d-4}-6l^{-2}\Lambda R^{d-2} 
    +2(d-2)\left(\Lambda^{-1}\right)'R'R^{d-3}\right. \nonumber \\ 
    && \left.+\,B N\Lambda^{-1}\left(R'\right)^2R^{d-4}+ 
    2(d-2)N\Lambda^{-1}R''R^{d-3}\right)\,, 
    \label{constraints_n}\\ 
    H_r &=& P_R R' -\Lambda P_\Lambda'-\Gamma P_\Gamma'\,, 
    \label{constraints_nr}\\ 
    G &=& -P_\Gamma'\,. 
    \label{constraints_g} 
\end{eqnarray} 

The equations of motion are obtained through the variation of 
the action (\ref{ham_action}), with the constraints defined in 
Eqs.\ (\ref{constraints_n})-(\ref{constraints_g}), i.e., 
\begin{eqnarray}\label{eom_1} 
  \dot{\Lambda} &=& (2\,a(d-2))^{-1}N\Lambda P_\Lambda R^{-(d-2)} 
  -(2\,a(d-2))^{-1}NP_RR^{-(d-3)}+(N^r\Lambda)'\,,\\ 
  \dot{R} &=& -(2\,a(d-2))^{-1}NP_\Lambda R^{-(d-3)}+N^rR'\,,\\ 
  \dot{\Gamma} &=& (8\pi\,a)^{-1}N\Lambda P_\Gamma R^{-(d-2)} 
  +(N^r\Gamma)'+\tilde{\Phi}'\,,\\ 
    \dot{P}_\Lambda &=& -\frac{d-3}{4\,a(d-2)}NP_\Lambda^2R^{-(d-2)} 
-\frac{1}{16\pi\,a}NP_\Gamma^2R^{-(d-2)}+a\,k\,NBR^{d-4}\nonumber\\ 
    && 
+6l^{-2}\,a\,NR^{d-2}-\left[2\,a(d-2)NR'R^{d-3}\right]'\Lambda^{-2} 
    +a\,NB\left(R'\right)^2\Lambda^{-2}R^{d-4}\nonumber\\ 
    && +2\,a(d-2)NR''\Lambda^{-2}R^{d-3}+N^rP_\Lambda'\,,\\ 
    \dot{P}_R &=& \frac{d-3}{4\,a}N\Lambda P_\Lambda^2R^{-(d-1)}+ 
    \frac{d-2}{16\pi\,a}N\Lambda P_\Gamma^2R^{-(d-1)} 
    -\frac{d-3}{2\,a(d-2)}NP_RP_\Lambda R^{-(d-2)}\nonumber\\ 
    &&+a\,k\,(d-4)\,BN\Lambda R^{d-5}+6l^{-2}\,a(d-2)N\Lambda R^{d-3}+ 
    \left[2\,a\,BN\left(\Lambda^{-1}\right)'\right]'R^{d-3}\nonumber\\ 
    && -a\,(d-4)BN\Lambda^{-1}(R')^2R^{d-5}+\left[2\,a\,BN\Lambda^{-1} 
    R^{d-4}\right]'R'\nonumber\\ 
    &&-\left(2\,a(d-2)N\Lambda^{-1}R^{d-3}\right)''+(N^rP_R)'\,,\\ 
    \dot{P}_\Gamma &=& N^rP_\Gamma'\,. 
    \label{eom_6} 
\end{eqnarray} 
In order to have a well defined variational principle, 
with which to derive the equations of motion above 
(\ref{eom_1})-(\ref{eom_6}), we need to eliminate the 
surface terms which remain from the variation of the 
original bulk action. 
These surface terms are eliminated through judicious 
choice of extra surface terms which should be added to 
the action (\ref{ham_action}). 
The action (\ref{ham_action}) has the 
following extra surface terms, after variation 
\begin{eqnarray} 
\textrm{Surface terms}&=& 2\,aN\Lambda^{-2} 
\left(R^{d-2}\right)\delta\Lambda +2\,a(d-2)N'\Lambda^{-1} 
R^{d-3}\delta R-2\,a(d-2)N\Lambda^{-1}R^{d-3}\delta R'\nonumber\\ 
&&-N^rP_R\delta R+N^r\Lambda\delta P_\Lambda+N^r\Gamma\delta 
P_\Gamma+\tilde{\Phi}\delta P_\Gamma\,. 
\label{surface_terms} 
\end{eqnarray} 
In order to evaluate this expression, we need to know the asymptotic 
conditions of the above functions individually. 

Starting with the limit $r\rightarrow 0$, we assume 
\begin{eqnarray} 
\label{falloff_0_i} 
\Lambda(t,r) &=& \Lambda_0(t)+O(r^2)\,,\\ 
R(t,r)&=& R_0(t)+R_2(t)r^2+O(r^4)\,,\\ 
P_\Lambda(t,r) &=& O(r^3)\,,\\ 
P_R(t,r) &=& O(r)\,,\\ 
N(t,r) &=& N_1(t)r+O(r^3)\,,\\ 
N^r(t,r) &=& N^r_1(t)r+O(r^3)\,,\\ 
\Gamma(t,r) &=& O(r)\,,\\ 
P_{\Gamma}(t,r) &=& X_0(t) + X_2(t) r^2 + O(r^4)\,,\\ 
\tilde{\Phi}(t,r)&=& \tilde{\Phi}_0 (t) + O(r^2)\,. 
 \label{falloff_0_f} 
\end{eqnarray} 
It is useful to redefine 
$X_{0}$ and $X_{2}$ in terms of two new quantities 
$Q_{0}$ and $Q_{2}$, such that $X_{0}=K_d\,{Q}_{0}$ 
and $X_{2}=K_d\,{Q}_{2}$, with 
$K_d\equiv8\pi a\sqrt{(d-2)(d-3)/2\pi}$. 
The surface terms for $r\to0$ suffer a modification 
in the definition $P_\Gamma$, in relation 
to the $d=4$ case \cite{louko3}. 

For $r\to\infty$ the fall-off conditions have to deal with the AdS 
asymptotic properties in $d$ dimensions. In $d=4$, the results in 
\cite{ht} were the basis for the $r\to\infty$ fall-off adopted in 
\cite{louko3}.  Since one now knows the Kerr-AdS solution 
in $d$ dimensions \cite{kerrantidesitter} one could 
follow \cite{ht} to find the fall-off conditions for our case. 
Here we take a more pragmatic approach, by simply 
generalizing, with some care, to $d$ 
dimensions the results in  \cite{louko3} for $d=4$. 
Hence, we assume the following conditions: 
\begin{eqnarray}\label{falloff_inf_i} 
    \Lambda(t,r) &=& f(l\,r^{-1})+l^3\lambda(t)r^{-d}+ 
    O^\infty(r^{-(d+1)})\,,\\ 
    R(t,r) &=& r+l^2\rho(t)r^{-(d-2)}+ 
    O^\infty(r^{-(d-1)})\,,\\ 
    P_\Lambda(t,r) &=& O^\infty(r^{-(d-2)})\,,\\ 
    P_R(t,r) &=& O^\infty(r^{-d})\,,\\ 
    N(t,r) &=& R'\Lambda^{-1}\left[N_+(t)+ 
    O^\infty(r^{-(d+1)})\right]\,,\\ 
    N^r(t,r) &=& O^\infty(r^{-(d-2)})\,,\\ 
    \Gamma(t,r)&=&O^\infty(r^{-(d-2)})\,,\\ 
    P_\Gamma(t,r)&=&X_+(t)+O^\infty(r^{-(d-3)})\,,\\ 
    \tilde{\Phi}(t,r)&=&\tilde{\Phi}_+(t)+ 
    O^\infty(r^{-(d-3)})\,, 
    \label{falloff_inf_f} 
\end{eqnarray} 
where we defined the function $f(l\,r^{-1})$ as 
\begin{eqnarray} 
    f(l\,r^{-1}):=l\,r^{-1} 
    \left(1+\sum_{s=1}^{[\frac{d-1}{2}]}(-1)^s\, 
    \frac{(2s-1)!!}{2^s\cdot 
s!}\left(l\,r^{-1}\right)^{2s}\,k^s\right)\,, 
    \label{f_function} 
\end{eqnarray} 
with the double factorial being defined by
\begin{eqnarray} 
s!!&\equiv& \left\{ \begin{array}{cc} s\cdot (s-2) 
\ldots 5\cdot3\cdot1 & s>0 \,\, \textrm{odd}\\ s\cdot (s-2)\ldots 
6\cdot4\cdot2 & s>0 \,\, \textrm{even}\\ 1 & s=-1,0 
\end{array}\right.\,. 
\label{doublefactorial} 
\end{eqnarray} 
where $[x]$ is the integer part of a given $x\in\mathbb{Z}/2$. 
Again, $l$ is the AdS length, and $k$ yields the topology of the horizon. 
As in $r\to0$ case, it is also useful 
to define $X_+\equiv K_d\,{\bar Q}_+$, with $K_d$ 
defined above. 

The surfaces terms which must be added to (\ref{ham_action}) 
in order to obtain a well defined variational principle 
are then 
\begin{eqnarray} 
S_{\partial\Sigma}\left[\Lambda,\,R,\,X_0,\,X_+; 
N,\,\tilde{\Phi}_0,\,\tilde{\Phi}_+\right]&=& 
\int\,dt\,\left(2\,aN_1\Lambda_0^{-1}R_0^{d-2}-N_+M_++ 
\tilde{\Phi}_0X_0-\tilde{\Phi}_+X_+\right)\,, 
\label{surface_action} 
\end{eqnarray} 
where $M_+(t)$ is defined as 
\begin{eqnarray} 
    M_+(t)\equiv2\,a(d-2) 
    \left(\lambda(t)-(d-1)\rho(t)\right)\,. 
    \label{m_plus} 
\end{eqnarray} 
One also should impose the condition of fixing $N_1\Lambda_0^{-1}$ and 
$\tilde{\Phi}_0$ on the horizon $r\to0$, which is the same as saying 
that $\delta(N_1\Lambda_0^{-1})=0$ and $\delta\tilde{\Phi}_0=0$.  The 
first condition fixes the rate of the boost suffered by the future 
unit normal to the constant $t$ hypersurfaces at the bifurcation 
horizon and is important to have a well defined Hamiltonian 
thermodynamics.  The second condition fixes the electric potential at 
the horizon.  At infinity, $r\to\infty$, one fixes $N_+$ and 
$\tilde{\Phi}_+$, which means $\delta N_+=0$ and 
$\delta\tilde{\Phi}_+=0$, in order that the variational principle is 
completely defined. 
So the total action is then the sum of the integral 
in Eq.\  (\ref{surface_action}) and the integral 
Eq.\  (\ref{ham_action}), 
\begin{eqnarray} 
S_{\Sigma+\partial\Sigma}\left[\Lambda,\,R,\,\Gamma,\, 
P_\Lambda,\,P_R,\,P_\Gamma;N,\,N^r,\,\tilde{\Phi}\right]&=& 
S_\Sigma\left[\Lambda,\,R,\,\Gamma,\,P_\Lambda,\,P_R,\,P_\Gamma; 
N,\,N^r,\,\tilde{\Phi}\right]\nonumber\\ 
&& + S_{\partial\Sigma}\left[\Lambda,\,R,\,X_0,\,X_+; 
N,\,\tilde{\Phi}_0,\,\tilde{\Phi}_+\right]\,. 
\label{total_action} 
\end{eqnarray} 
With this action one then derives the equations of motion 
(\ref{eom_1})-(\ref{eom_6}) without extra surface terms, 
provided that the fixing of $N_1\Lambda_0^{-1}$, 
$\tilde{\Phi}_0$, $N_+$, and $\tilde{\Phi}_+$ is 
taken into account 
\cite{louko1,louko3,louko4,kunst1,kunst2,bose,louko2,dl3,dl4,louko5}. 

\subsection{Reconstruction, canonical transformation, and action} 
\label{reconstruction} 
In order to reconstruct the mass and the time from the canonical data, 
which amounts to making a canonical transformation, we have to rewrite 
the form of the solutions of Eqs.\ (\ref{general_metric}). 
In the process, we have to consider how to reconstruct the charge from 
the canonical data, which is on the hypersurface embedded in this 
charged spacetime.  We follow Kucha\v{r} \cite{kuchar} for this 
reconstruction.  We concentrate our analysis on the right static 
region of the Carter-Penrose diagram. 

Developing the Killing time $T$ as function of $(t,r)$ in 
the expression for the vector potential (\ref{vector_potential}), 
and making use of the gauge freedom that allows us to write 
\begin{eqnarray} 
    A &=& \sqrt{\frac{d-2}{8\pi(d-3)}}\frac{\bar{Q}}{R^{d-3}}\,dT+d\xi\,, 
    \label{vector_potential_plusgauge} 
\end{eqnarray} 
 where $\xi(t,r)$ is an arbitrary continuous function of $t$ and $r$, 
 we can write the one-form potential as 
\begin{eqnarray} 
    A &=& \left(\sqrt{\frac{d-2}{8\pi(d-3)}}\frac{\bar{Q}}{R^{d-3}}T'+ 
    \xi'\right)dr+ 
    \left(\sqrt{\frac{d-2}{8\pi(d-3)}}\frac{\bar{Q}}{R^{d-3}}\dot{T}+ 
    \dot{\xi}\right)dt\,. 
    \label{vector_potential_plusgauge_tr} 
\end{eqnarray} 
Equating the expression (\ref{vector_potential_plusgauge_tr}) 
with Eq.\  (\ref{a_ansatz}), and making use of the definition 
of $P_{\Gamma}$ in Eq.\  (\ref{p_gamma}), we arrive at 
\begin{eqnarray} 
    P_{\Gamma} &=& K_d\, \bar{Q}\,. 
    \label{pgamma_kd_q} 
\end{eqnarray} 
with $K_d\equiv4\,a\sqrt{2\pi\,(d-2)(d-3)}$. 
We also obtain the derivative of the gauge function with 
respect to $r$ 
\begin{eqnarray} 
    \xi'&=&\Gamma\,+\,K_d^{-2}R^{-2(d-3)}P_\Gamma 
    F^{-1}\Lambda P_\Lambda\,. 
    \label{xi_derivado_r} 
\end{eqnarray} 
In the static region, we have defined $F$ as 
\begin{equation} 
F(t,r) = k\,+l^{-2} R^2-\frac{2\,\bar{M}}{R^{d-3}}+ 
\frac{\bar{Q}^2}{R^{2(d-3)}}\,. 
\label{f_1} 
\end{equation} 
We now make the following substitutions 
\begin{equation} 
T=T(t,r)\,, \qquad \qquad R=R(t,r)\,, 
\end{equation} 
into the solution  (\ref{general_metric}), getting 
\begin{eqnarray} 
ds^2 &=& -(F\dot{T}^2-F^{-1}\dot{R}^2)\,dt^2\,+ 
\,2(-FT'\dot{T}+F^{-1}R'\dot{R})\,dtdr\,+\,(-F(T')^2+F^{-1}\dot{R}^2)\,dr^2\,+ 
\,R^2\,(d\,\Omega_{d-2}^k)^2\,. 
\end{eqnarray} 
This introduces the ADM foliation directly into the solutions. 
Comparing it with the ADM metric (\ref{adm_ansatz}), written in 
another form as 
\begin{equation} 
ds^2 = 
-(N^2-\Lambda^2(N^r)^2)\,dt^2\,+\,2\Lambda^2N^r\,dtdr\,+ 
\,\Lambda^2dr^2\,+\,R^2\,(d\,\Omega_{d-2}^k)^2\,, 
\end{equation} 
we can write a set of three equations 
\begin{eqnarray} 
\Lambda^2 &=& -F(T')^2+F^{-1}(R')^2\,,\label{adm_1}\\ 
\Lambda^2N^r &=& -FT'\dot{T}+F^{-1}R'\dot{R}\,, \label{adm_2}\\ 
N^2-\Lambda^2(N^r)^2 &=& F\dot{T}^2-F^{-1}\dot{R}^2\,. \label{adm_3} 
\end{eqnarray} 
The first two equations, Eqs.\ (\ref{adm_1}) and Eq.\  (\ref{adm_2}), 
give 
\begin{equation} \label{shift_def} 
N^r = \frac{-FT'\dot{T}+F^{-1}R'\dot{R}}{-F(T')^2+F^{-1}(R')^2}\,. 
\end{equation} 
This one solution, together  with Eq.\  (\ref{adm_1}), give 
\begin{equation} \label{lapse_def} 
N = \frac{R'\dot{T}-T'\dot{R}}{\sqrt{-F(T')^2+F^{-1}(R')^2}}\,. 
\end{equation} 
One can show that $N(t,r)$ is positive (see \cite{kuchar}). 
Next, putting Eqs.\  (\ref{shift_def})-(\ref{lapse_def}), 
into the definition of the conjugate momentum of 
the canonical coordinate $\Lambda$, given in 
Eq.\  (\ref{p_lambda}), one finds 
the spatial derivative of $T(t,r)$ as a function of the canonical 
coordinates, i.e., 
\begin{equation} 
\label{t_linha} 
-T' = (2\,a(d-2))^{-1}R^{-(d-3)}F^{-1}\Lambda P_\Lambda\,. 
\end{equation} 
Later we will see that $-T'=P_M$, as it 
will be conjugate to a new canonical coordinate $M$, 
defined below in Eq.\  (\ref{m_pequena}). 
Following this procedure to the end, we may then find the form of the 
new coordinate $\bar{M}(t,r)$, as a function of $t$ and $r$. First, we 
need to know the form of $F$ as a function of the canonical pair 
$\Lambda\,,\,R$. For that, we replace back into Eq.\  (\ref{adm_1}) the 
definition, in Eq.\  (\ref{t_linha}), of $T'$, giving 
\begin{equation} 
\label{f_2} 
F = \left(\frac{R'}{\Lambda}\right)^2- 
\left(\frac{P_\Lambda}{2\,a(d-2)R^{d-3}}\right)^2\,. 
\end{equation} 
Equating this form of $F$ with Eq.\  (\ref{f_1}), we obtain 
\begin{equation} 
\label{m_grande} 
\bar{M}(t,r) = \frac12 R^{d-3}\left(k+l^{-2}R^2+ 
\frac{P_\Gamma^2K_d^{-2}}{R^{2(d-3)}}-F\right)\,, 
\end{equation} 
where $F$ is given in Eq.\  (\ref{f_2}). 
However, it will be more convenient for what is coming 
to define the following mass 
\begin{eqnarray} 
    M(t,r)&\equiv& 2\,a\,(d-2)\bar{M}(t,r)\,, 
    \label{m_pequena} 
\end{eqnarray} 
where $\bar{M}(t,r)$ is defined in Eq.\  (\ref{m_grande}). 
The new canonical coordinate is thus $M$, which 
can also be called the canonical mass. It 
is now a straightforward calculation to determine the Poisson bracket 
of this variable with $P_M=-T'$ and see that they are conjugate, thus 
making Eq.\  (\ref{t_linha}) the conjugate momentum of $M$, i.e., 
\begin{equation}\label{p_m} 
P_M = (2\,a(d-2))^{-1}R^{-(d-3)}F^{-1}\Lambda P_\Lambda\,. 
\end{equation} 
In the same fashion, one needs to find the 
charge canonical coordinate related to $\bar{Q}$. In this case 
the relation is trivial, and one finds that the 
canonical charge coordinate $Q$, or simply 
the canonical charge, is given by 
\begin{eqnarray} 
Q &=& \bar{Q}\,. 
\label{qqbar} 
\end{eqnarray} 
Then, the other natural transformation is the one that 
relates $P_{\Gamma}$ with the canonical charge $Q$. 
Through the relation of $P_{\Gamma}$ with the charge parameter $\bar{Q}$, 
$P_{\Gamma}=K_d \bar{Q}$, the canonical transformation between the old 
variable $P_{\Gamma}$ and the new canonical variable $Q$ is then 
immediately found to be 
$P_{\Gamma}=K_d Q$. 
Like $M$ previously, $Q$ is  a natural choice for the canonical 
coordinate, 
because it is equal to the charge parameter of 
the metric solution in Eq.\  (\ref{general_metric}). 
The one coordinate which remains to be found is $P_{Q}$, 
the conjugate momentum to the charge. 
It is also necessary to find out the other new canonical 
variable which commutes with $M$, $P_M$, and $Q$, 
and which guarantees, with its conjugate 
momentum, that the transformation from $\Lambda,\,R,\,\Gamma$, 
to $M$, $Q$, and the new variable is canonical. 
Immediately is it seen that $R$ commutes 
with $M$, $P_M$, and $Q$. It is then a candidate. It remains to be 
seen 
whether $P_R$ also commutes with $M$, $P_M$, and $Q$.  As with $R$, 
it is 
straightforward to see that $P_R$ does not commute with $M$ and $P_M$, 
as these contain powers of $R$ in their definitions, and 
$\left\{R(t,r),\,P_R(t,r^*) \right\}=\delta(r-r^*)$.  So, rename the 
canonical variable $R$ as $R=\textrm{R}$.  We have then to find a new 
conjugate momentum to $\textrm{R}$ which also commutes with $M$, $P_M$, 
and $Q$, making the transformation from 
$\left\{\Lambda,\,R,\,\Gamma;\,P_\Lambda,\,P_R,\,P_{\Gamma}\,\right\} 
\rightarrow\, 
\left\{M,\,\textrm{R},\,Q;\,P_M,\,P_{\textrm{R}},\,P_{Q}\,\right\}$ 
a 
canonical one. 
The way to proceed is to look at the constraint $H_r$, which is called 
in this formalism the super-momentum. This is the constraint which 
generates spatial diffeomorphisms in all variables. Its form, in the 
initial canonical coordinates, is 
$H_r=-\Lambda\,P_\Lambda'+P_R\,R'-\Gamma P_{\Gamma}'$. 
In this formulation, $\Lambda$ is a spatial density, $R$ is a 
spatial scalar, and $\Gamma$ is also a spatial density. 
As the new variables, $M$, $\textrm{R}$, and $Q$, are 
spatial scalars, the generator of spatial diffeomorphisms is written 
as $H_r=P_M M'+P_{\textrm{R}} \textrm{R}'+P_{Q} Q'$, 
regardless of the 
particular form of the canonical coordinate transformation. It is thus 
equating these two expressions of the super-momentum $H_r$, with $M$, 
$P_M$, and $Q$  written as functions of $\Lambda,\,R,\,\Gamma$ 
and their respective 
momenta, that gives us one equation for the new $P_{\textrm{R}}$ and 
$P_{Q}$. This means that we have two unknowns, $P_{\textrm{R}}$ and 
$P_{Q}$, for one equation only. This suggests that we should make 
the coefficients of $R'$ and $P_{\Gamma}'$ equal to zero independently. 
This results in 
\begin{eqnarray} 
  P_{\textrm{R}}&=&P_R-\frac12(d-3)\,k\,R^{-1}F^{-1}\Lambda 
  P_\Lambda-\frac12(d-1)l^{-2}RF^{-1}\Lambda P_\Lambda+ 
  \frac12(d-3)K_d^{-2}P_\Gamma^2R^{-2d+5}F^{-1}\Lambda 
  P_\Lambda\nonumber\\ 
  && -\frac12R^{-1}\Lambda P_\Lambda+R''\Lambda^{-1}F^{-1} 
  P_\Lambda-\Lambda'R'F^{-1}\Lambda^{-2}P_\Lambda- 
  P_\Lambda'F^{-1}\Lambda^{-1}\nonumber\\ 
  &&+(d-3)R'R^{-1}F^{-1}\Lambda^{-1}P_\Lambda\,, 
  \label{p_rnew}\\ 
  P_{Q} &=& -K_d\,\Gamma-K_d^{-1}R^{-2(d-3)}P_\Gamma 
  F^{-1}\Lambda P_\Lambda\,. 
  \label{p_q} 
\end{eqnarray} 
(Note that in addition one finds $P_{Q}=-K_d\,\xi'$.) 
It can then be shown that $P_{Q}$ commutes with the new $P_{\textrm{R}}$ 
and with the rest of the new coordinates, except with $Q$. 
We have now all the canonical variables of the new set determined. For 
completeness and future use, we write the inverse transformation for 
$\Lambda$ and $P_\Lambda$, 
\begin{eqnarray} \label{inversetrans_1} 
\Lambda &=& \left((\textrm{R}')^2F^{-1}-P_M^2F\right)^{\frac12}\,, \\ 
P_\Lambda &=& 2\,a(d-2)R^{d-3}FP_M 
\left((\textrm{R}')^2F^{-1}-P_M^2F\right)^{-\frac12}\,. 
\label{inversetrans_2} 
\end{eqnarray} 
In summary, the full set of canonical transformations are the following, 
\begin{eqnarray} 
  \label{setcanonicaltrans} 
  \textrm{R} &=& R\,,\nonumber \\ 
  M &=& a(d-2) R^{d-3}\left(k+l^{-2}R^2+ 
  \frac{P_\Gamma^2K_d^{-2}}{R^{2(d-3)}}-F\right)\,, 
  \nonumber \\ 
  Q &=& P_{\Gamma}K_d^{-1}\,, \nonumber \\ 
  P_{\textrm{R}} &=& P_R-\frac12(d-3)\,k\,R^{-1}F^{-1}\Lambda 
  P_\Lambda-\frac12(d-1)l^{-2}RF^{-1}\Lambda P_\Lambda+ 
  \frac12(d-3)K_d^{-2}P_\Gamma^2R^{-2d+5}F^{-1}\Lambda 
  P_\Lambda\nonumber\\ 
  && -\frac12R^{-1}\Lambda P_\Lambda+R''\Lambda^{-1}F^{-1} 
  P_\Lambda-\Lambda'R'F^{-1}\Lambda^{-2}P_\Lambda- 
  P_\Lambda'F^{-1}\Lambda^{-1}\nonumber\\ 
  &&+(d-3)R'R^{-1}F^{-1}\Lambda^{-1}P_\Lambda\,, 
  \nonumber \\ 
  P_M &=& (2\,a(d-2))^{-1}R^{-(d-3)}F^{-1}\Lambda P_\Lambda\,, 
  \nonumber \\ 
  P_{Q} &=& -K_d\,\Gamma-K_d^{-1}R^{-2(d-3)}P_\Gamma 
  F^{-1}\Lambda P_\Lambda\,. 
\end{eqnarray} 
It remains to be seen that this set of transformations 
is in fact canonical. 
In order to prove that the set of equalities in expression 
(\ref{setcanonicaltrans}) is canonical we start with the equality 
\begin{eqnarray} 
  P_\Lambda \delta\Lambda + P_R\delta R +P_\Gamma \delta\Gamma 
  &-&P_M \delta M - P_{\textrm{R}}\delta\textrm{R}-P_{Q}\delta 
Q 
  = \left(a(d-2)R^{d-3}\delta R\ln 
  \left|\frac{2\,a(d-2)R^{d-3}R'+\Lambda P_\Lambda} 
  {2a(d-2)R^{d-3}R'-\Lambda P_\Lambda}\right|\right)'+ 
  \nonumber \\ 
  &+&\delta\left(\Gamma P_{\Gamma}+\Lambda P_\Lambda + 
  a(d-2)R^{d-3}R'\ln\left|\frac{2\,a(d-2)R^{d-3}R'-\Lambda P_\Lambda} 
  {2\,a(d-2)R^{d-3}R'+\Lambda P_\Lambda}\right|\right)\,. 
  \label{oddidentity} 
\end{eqnarray} 
We now integrate expression (\ref{oddidentity}) in $r$, in 
the interval from $r=0$ to $r=\infty$. The first term on the right 
hand side of Eq.\  (\ref{oddidentity}) vanishes due to the falloff 
conditions (see Eqs.\ (\ref{falloff_0_i})-(\ref{falloff_0_f}) and 
Eqs.\ (\ref{falloff_inf_i})-(\ref{falloff_inf_f})). 
We then obtain the following expression 
\begin{eqnarray} \label{int_oddidentity} 
\int_0^\infty\,dr\,\left(P_\Lambda \delta\Lambda + P_R \delta 
  R + P_\Gamma \delta\Gamma \right) 
  -\int_0^\infty\,dr\,\left(P_M \delta M + 
P_{\textrm{R}}\delta\textrm{R} + P_{Q} \delta Q \right) &=& 
\delta\omega\,\left[\Lambda,\,R,\,\Gamma,\,P_\Lambda,\,P_\Gamma\right]\,, 
\end{eqnarray} 
where 
$\delta\omega\,\left[\Lambda,\,R,\,\Gamma,\,P_\Lambda,\,P_\Gamma\right]$ 
is a well defined functional, which is also an exact form. This equality 
shows 
that the difference between the Liouville form of 
$\left\{R,\,\Lambda,\,\Gamma;\,P_R,\,P_\Lambda,\,P_\Gamma\right\}$ 
and the Liouville form 
of 
$\left\{\textrm{R},\,M,\,Q;\,P_{\textrm{R}},\,P_M,\,P_{Q}\right\}$ 
is an 
exact 
form, which implies that the transformation of variables given by the 
set of equations (\ref{setcanonicaltrans}) is canonical. 

Armed with the certainty of the canonicity of the new variables, we 
can write the asymptotic form of the canonical variables and of the 
metric function $F(t,r)$. These are, for $r\rightarrow 0$ 
\begin{eqnarray} \label{newfalloff_0_i} 
F(t,r) &=& 4 R_2^2(t) \Lambda_0^{-2}(t) r^2 + O(r^4)\,, \\ 
\textrm{R}(t,r) &=& R_0(t)+R_2(t)\,r^2+O(r^4)\,, \\ 
M(t,r) &=& M_0(t) + M_2(t) \, r^2+O(r^4)\,, \\ 
Q(t,r) &=& Q_0(t) + Q_2(t) \, r^2+O(r^4)\,, \\ 
P_\textrm{R}(t,r) &=& O(r)\,, \\ 
P_M(t,r) &=& O(r)\,,\\ 
P_{Q}(t,r) &=& O(r)\,. 
\label{newfalloff_0_f} 
\end{eqnarray} 
with 
\begin{eqnarray} 
    M_0 &=& a(d-2)R_0^{d-3}\left(l^{-2}R_0^2+\,k\,+P_\Gamma^2 
    K_d^{-2}R_0^{-2(d-3)}\right)\,,\\ 
    M_2 &=& a(d-2)\left((d-3)R_0^{d-4} 
    \left[l^{-2}R_0^2+\,k\,+\frac{Q_0^2}{R_0^{2(d-3)}}\right]R_2 
    \right.\nonumber\\ 
    &&\left.+R_0^{d-3}\left[\frac{2Q_0Q_2}{R_0^{2(d-3)}}- 
    \frac{2(d-3)Q_0^2R_2}{R_0^{2d-5}}+2l^{-2}R_0R_2- 
    4R_2\Lambda_0^{-2}\right]\right)\,. 
\end{eqnarray} 
For $r\rightarrow \infty$, we have 
\begin{eqnarray} \label{newfalloff_inf_i} 
\textrm{R}(t,r) &=& r+l^2\rho(t)r^{-(d-2)}+ 
    O^\infty(r^{-(d-1)})\,, \\ 
M(t,r) &=& M_+(t) + O^{\infty}(r^{-(d-3)})\,, \\ 
Q(t,r) &=& Q_+(t) + O^{\infty}(r^{-(d-3)})\,, \\ 
P_\textrm{R}(t,r) &=& O^\infty(r^{-d})\,, \\ 
P_M(t,r) &=& O^\infty(r^{-(d+2)})\,, \\ 
P_{Q}(t,r) &=& O^{\infty}(r^{-(d-2)})\,. 
\label{newfalloff_inf_f} 
\end{eqnarray} 
where $M_+(t)=2\,a(d-2)\left(\lambda(t)-(d-1)\rho(t)\right)$, 
as seen before in Eq.\  (\ref{m_plus}) and Eqs. 
(\ref{m_grande})-(\ref{m_pequena}). 

We are now almost ready to write the action with the new canonical 
variables. It is now necessary to determine the new Lagrange 
multipliers. In order to write the new constraints with the new 
Lagrange multipliers, we can use the identity given by the space 
derivative of $M$, 
\begin{equation} 
M' = -\Lambda^{-1}\left(R'H+(2\,a(d-2))^{-1}R^{-(d-3)}P_\Lambda 
\left(H_r-\Gamma G\right)\right) 
+2\,a(d-2)\,K_d^2P_\Gamma P_\Gamma'R^{-(d-3)}\,. 
\end{equation} 
Solving for $H$ and making use of the inverse transformations of 
$\Lambda$ and $P_\Lambda$, in Eqs.\ (\ref{inversetrans_1}) and 
(\ref{inversetrans_2}), we get 
\begin{eqnarray} \label{oldconstraintsinnewvariables_1} 
H &=& - \frac{M'F^{-1}\textrm{R}'+FP_MP_{\textrm{R}}+ 
2\,a(d-2)R^{-(d-3)}QQ'R'F^{-1}} 
{\left(F^{-1}(\textrm{R}')^2-FP_M^2\right)^{\frac12}}\,, \\ 
H_r &=& P_M M' + P_{\textrm{R}} 
\textrm{R}'+P_{Q} Q'\,, \\ 
G &=& -K_d\,Q'\,. \label{oldconstraintsinnewvariables_2} 
\end{eqnarray} 
Following Kucha\v{r} \cite{kuchar}, the new set of constraints, totally 
equivalent to the old set $H(t,r)=0$, $H_r(t,r)=0$, and $G=0$, outside the 
horizon points, is $M'(t,r)=0$, $P_{\textrm{R}}(t,r)=0$, and 
$Q'(t,r)=0$. 
By continuity, this also applies on the horizon, where $F(t,r)=0$. 
So we can say that the equivalence is valid everywhere. 

The new Hamiltonian, which is the total sum of the constraints, 
can now be written as 
\begin{equation} \label{newHamiltonian} 
NH+N^rH_r+\tilde{\Phi}G= N^M M' + N^{\textrm{R}} 
P_{\textrm{R}}+N^{Q}Q'\,. 
\end{equation} 
In order to determine the new Lagrange multipliers, one has to write 
the left hand side of the previous equation, 
Eq.\  (\ref{newHamiltonian}), and replace the constraints on that side 
by their expressions as functions of the new canonical coordinates, 
spelt out in Eqs. 
(\ref{oldconstraintsinnewvariables_1})-\noindent 
(\ref{oldconstraintsinnewvariables_2}). 
After manipulation, one gets 
\begin{eqnarray} \label{mult_trans_1} 
  N^M &=& - NF^{-1}R'\Lambda^{-1}+ 
  (2\,a(d-2))^{-1}N^rR^{-(d-3)}F^{-1}\Lambda P_\Lambda\,, \\ 
  N^{\textrm{R}} &=& -(2\,a(d-2))^{-1}NR^{-(d-3)}P_\Lambda + N^r 
  R'\,,\label{mult_trans_2}\\ 
  N^{Q} &=& 2\,a(d-2)NR^{-(d-3)}\Lambda^{-1}K_d^{-1}P_\Gamma 
  F^{-1}R'-N^rK_d\,\Gamma-N^rK_d^{-1}R^{-2(d-3)}P_\Gamma 
  F^{-1}\Lambda P_\Lambda 
  \,,\label{mult_trans_3} 
\end{eqnarray} 
allowing us determine its asymptotic conditions from the original 
conditions given above. 
These transformations are non-singular for $r>0$. 
As before, for $r\rightarrow 0$, 
\begin{eqnarray} 
  \label{mult_newfalloff_0_i} 
  N^M(t,r) &=& -\frac12 N_1(t) \Lambda_0 R_2^{-1}+O(r^2)\,,\\ 
  N^{\textrm{R}}(t,r) &=& O(r^2)\,,\\ 
  N^{Q}(t,r) &=& -K_d\tilde{\Phi}_0(t)+ 
  a(d-2)N_1Q_0\Lambda_0R_2^{-1}R_0^{-(d-3)}+O(r^2)\,, 
  \label{mult_newfalloff_0_f} 
\end{eqnarray} 
and for $r\rightarrow\infty$ we have 
\begin{eqnarray} 
  \label{mult_newfalloff_inf_i} 
  N^M(t,r) &=& -\tilde{N}_+(t) +  O^\infty(r^{-(d+1)})\,,\\ 
  N^{\textrm{R}}(t,r) &=& O^\infty(r^{-(d-2)})\,,\\ 
  N^{Q}(t,r)&=&-K_d\tilde{\Phi}_+(t)+O^\infty(r^{-(d-3)})\,. 
  \label{mult_newfalloff_inf_f} 
\end{eqnarray} 
The conditions 
(\ref{mult_newfalloff_0_i})-(\ref{mult_newfalloff_inf_f}) 
show that the transformations in 
Eqs.\ (\ref{mult_trans_1})-(\ref{mult_trans_2}) are satisfactory in 
the case of $r\rightarrow\infty$, but not for $r\rightarrow 0$. This 
is due to fact that in order to fix the Lagrange multipliers for 
$r\rightarrow\infty$, as we are free to do, we fix $\tilde{N}_+(t)$, 
which we already do when adding the surface term 
$ 
- \int\, dt \, \tilde{N}_+ M_+ 
$ 
to the action, in order to obtain the equations of motion in the bulk, 
without surface terms. 
The same is true for $\tilde{\Phi}_+$. 
However, at $r=0$, we see that fixing the multiplier $N^M$ to values 
independent of the canonical variables is not the same as fixing $N_1 
\Lambda_0^{-1}$ to values independent of the canonical variables. 
The same is true of the fixation of $N^{Q}$ with respect to 
$\tilde{\Phi}_0$. 
We need to rewrite the multipliers $N^M$ and $N^{Q}$ for the 
asymptotic 
regime 
$r\rightarrow 0$ without affecting their behavior for 
$r\rightarrow\infty$. 
In order to proceed we have to make one assumption, which is that the 
expression given in asymptotic condition of $M(t,r)$, as $r\rightarrow 
0$, for the term of order zero $M_0(R_0,Q_0)$, 
defines $R_0$ as a function of $M_0$ and $Q_0$, 
and $R_0$ is the horizon radius function, $R_0\equiv 
R_{\textrm{h}}(M_0,Q_0)$. 
Also, we assume that $M_0>M_{\textrm{\tiny{crit}}}(Q_0)$, 
where $M_{\textrm{\tiny{\rm crit}}}(Q_0)$ 
is found through the system of two equations 
$R^{2(d-3)}F(R)=0$ and $(R^{2(d-3)}F(R))'=0$, with $F(R)$ defined 
in (\ref{general_f}). 
With these assumptions, we are working in the domain of the classical 
solutions. 
We can immediately obtain that the variation of 
$R_0$ is given in relation to the variations of $M_0$ and $Q_0$ as 
\begin{eqnarray} 
  \delta R_0 = \left\{a(d-2)\left((d-3)\,k\,R_0^{d-4}+ 
  (d-1)l^{-2}R_0^{d-2}-(d-3)Q_0^2R_0^{-(d-2)}\right)\right\}^{-1} 
  \left(\delta M_0-\frac{2\,a(d-2)Q_0}{R_0^{d-3}}\delta 
Q_0\right)\,. 
  \label{var_r_m} 
\end{eqnarray} 
This expression will be used when we derive the equations of 
motion from the new action. 
We now define the new multipliers $\tilde{N}^M$ and $\tilde{N}^{Q}$ 
as 
\begin{eqnarray} \label{new_n_m} 
  \tilde{N}^M &=& - N^M 
  \left[(1-g)+2R_0^{d-3}g\,\left((d-3)\,k\,R_0^{d-4}+ 
  (d-1)l^{-2}R_0^{d-2}-(d-3)Q_0^2R_0^{-(d-2)}\right)^{-1} 
  \right]^{-1}\,,\\ 
  \tilde{N}^{Q} &=& \tilde{N}^M 4\,g\,a\,(d-2)\,Q_0 
  \left((d-3)\,k\,R_0^{d-4}+(d-1)l^{-2}R_0^{d-2}-(d-3)Q_0^2 
  R_0^{-(d-2)}\right)^{-1} - N^{Q}\,, 
\label{new_n_q} 
\end{eqnarray} 
where $g(r)=1+O(r^2)$ for $r\rightarrow 0$ and $g(r)=O^\infty(r^{-5})$ 
for $r\rightarrow\infty$. The new multipliers, functions of the old 
multipliers $N^M$ and $N^{Q}$, have as their properties for 
$r\rightarrow0$, 
\begin{eqnarray} 
\tilde{N}^M(t,r) &=& \tilde{N}_0^M(t) + O(r^{2})\,,\\ 
\tilde{N}^{Q}(t,r) &=& K_d\,\tilde{\Phi}_0(t) + O(r^{2})\,, 
\end{eqnarray} 
and as their properties for $r\rightarrow \infty$, 
\begin{eqnarray} 
\tilde{N}^M(t,r) &=& \tilde{N}_+(t) + O^\infty(r^{-(d+1)})\,,\\ 
\tilde{N}^{Q}(t,r) &=& K_d\,\tilde{\Phi}_+(t) + 
O^\infty(r^{-(d-3)})\,. 
\end{eqnarray} 
When the constraints $M'=0=Q'$ hold, $\tilde{N}_0^M$ is given by 
\begin{equation} 
\tilde{N}_0^M =  N_1 \Lambda_0^{-1} \,. 
\end{equation} 
With this new constraint $\tilde{N}^M$, fixing $N_1 \Lambda_0^{-1}$ at 
$r=0$ or fixing $\tilde{N}_0^M$ is equivalent, there being no problems 
with $N^{\textrm{R}}$, which is left as determined in 
Eq.\  (\ref{mult_trans_2}). With respect to $\tilde{N}^{Q}$ the same 
happens, i. e., fixing the zero order term of the expansion of 
$N^{Q}$ for $r\to0$ is the same as fixing $\tilde{\Phi}_0$, even 
if multiplied by a dimension dependent constant $K_d$. 
At infinity there were no initial problems with the 
definitions of both $\tilde{N}^M$ and $\tilde{N}^{Q}$. 

The new action is now written as the sum of $S_\Sigma$, the bulk action, 
and 
$S_{\partial\Sigma}$, the surface action, 
\begin{eqnarray} 
  && S\left[M, \textrm{R}, Q, P_M, P_{\textrm{R}}, P_{Q}; 
\tilde{N}^M, 
  N^{\textrm{R}}, \tilde{N}^{Q}\right] = \nonumber \\ 
  && \,\,\,\,\qquad \int \,dt\, \int_0^\infty 
\, dr 
\, 
  \left\{ P_M\dot{M} + P_\textrm{R} \dot{\textrm{R}} + P_{Q} 
\dot{Q} 
  +\tilde{N}^{Q}Q' - N^{\textrm{R}}P_{\textrm{R}} + 
  \tilde{N}^M (1-g)\,M' 
  \right. \nonumber \\ 
  && \,\,\,\,\qquad \left. 
  +\tilde{N}^M\,2\,g\, 
  \left((d-3)\,k\,R_0^{d-4}+(d-1)l^{-2} 
  R_0^{d-2}-(d-3)Q_0^2R_0^{-(d-2)}\right)^{-1} 
  \right. \nonumber \\ 
  && \,\,\,\,\qquad \times\left. 
  \left(R_0^{d-3}\,M'-2\,a(d-2)Q'Q_0\right) 
  \right\} 
  \nonumber \\ 
  && \,\,\,\,\qquad + \int \, dt \, \left\{\left(2\,a\,\tilde{N}_0^M\, 
  R_0^{d-2}-\tilde{N}_+ M_+ \right) + 
  K_d\,\left(\tilde{\Phi}_0Q_0-\tilde{\Phi}_+Q_+\right) 
  \right\}\,. 
  \label{newaction} 
\end{eqnarray} 
The new equations of motion are now 
\begin{eqnarray} \label{new_eom_1} 
\dot{M} &=& 0\,, \\ 
\dot{\textrm{R}} &=& N^{\textrm{R}}\,, \\ 
\dot{Q} &=& 0\,,\\ 
\dot{P}_M &=& (N^M)'\,, \\ 
\dot{P}_{\textrm{R}} &=& 0\,, \\ 
\dot{P}_{Q} &=& (N^{Q})'\,,\\ 
M' &=& 0\,, \\ 
P_{\textrm{R}} &=& 0\,, \\ 
Q'&=&0\,,\label{new_eom_9} 
\end{eqnarray} 
where we understood $N^M$ to be a function of the new constraint, 
defined through Eq.\ (\ref{new_n_m}) and $N^{Q}$ as a function of the
new constraint defined through Eq.\ (\ref{new_n_q}).  The resulting
boundary terms of the variation of this new action, Eq.\
(\ref{newaction}), are, first, terms proportional to $\delta M$,
$\delta \textrm{R}$, and $\delta Q$ on the initial and final
hypersurfaces, and, second,
\begin{eqnarray}\label{surface_varied_terms_final} 
\int \, dt \, \left(2\,a R_0^{d-2} \delta\tilde{N}_0^M - 
M_+ \delta\tilde{N}_+ \right) + K_d\,\left(Q_0\delta\tilde{\Phi}_0- 
Q_+\delta\tilde{\Phi}_+\right) \,. 
\end{eqnarray} 
To arrive at (\ref{surface_varied_terms_final}) we have used 
the expression in Eq.\  (\ref{var_r_m}). 
The action in 
Eq.\  (\ref{newaction}) yields the equations of motion, 
Eqs.\ (\ref{new_eom_1})-(\ref{new_eom_9}), provided that we fix the 
initial and final values of the new canonical variables and that we 
also fix the values of $\tilde{N}^M_0$ and of $\tilde{N}_+$, and of 
$\tilde{\Phi}_0$ and $\tilde{\Phi}_+$. 
Thanks to 
the redefinition of the Lagrange multiplier, from $N^M$ to 
$\tilde{N}^M$, the fixation of those quantities, $\tilde{N}^M_0$ and 
$\tilde{N}_+$, has the same meaning it had before the 
canonical transformations and the redefinition of $N^M$. 
The same happens with $\tilde{\Phi}_0$ and $\tilde{\Phi}_+$. 
This keeping of meaning is guaranteed through the use of our gauge 
freedom to choose the multipliers, and at the same time not fixing 
the boundary variations independently of the choice of Lagrange 
multipliers, which in turn allow us to have a well defined 
variational principle for the action. 

\subsection{Hamiltonian reduction} 
\label{h_red} 
We now solve the constraints in order to reduce to the true dynamical 
degrees of freedom. The equations of motion 
(\ref{new_eom_1})-(\ref{new_eom_9}) allow us to write $M$ and $Q$ 
as independent functions of space, $r$, 
\begin{eqnarray} \label{m_t} 
M(t,r) &=& \textbf{m}(t)\,,\\ 
Q(t,r) &=& \textbf{q}(t)\,. 
\label{q_t} 
\end{eqnarray} 
The reduced action, with the constraints taken into account, is then 
\begin{equation} \label{red_action} 
S 
\left[\textbf{m},\textbf{p}_{\textbf{m}},\textbf{q},\textbf{p}_{\textbf{q}} 
;\tilde{N}_0^M,\tilde{N}_+,\tilde{\Phi}_0,\tilde{\Phi}_+\right] 
= \int dt\,\,\left(\textbf{p}_{\textbf{m}} 
\dot{\bf{m}}+\textbf{p}_{\textbf{q}} 
\dot{\bf{q}}-\textbf{h}\right)\,, 
\end{equation} 
where 
\begin{eqnarray} \label{new_p_m} 
\textbf{p}_{\textbf{m}} &=& \int_0^\infty dr\,P_M\,,\\ 
\textbf{p}_{\textbf{q}} &=& \int_0^\infty dr\,P_{Q}\,, 
\label{new_p_q} 
\end{eqnarray} 
and the reduced Hamiltonian, $\textbf{h}$, is now written as 
\begin{equation} \label{red_Hamiltonian} 
  \textbf{h}(\textbf{m},\,\textbf{q};t) = 
  -2\,a\,\tilde{N}_0^M R_{\textrm{h}}^{d-2}+ 
  \tilde{N}_+ \textbf{m}+ K_d\,\textbf{q} 
  \left(\tilde{\Phi}_+-\tilde{\Phi}_0\right)\,, 
\end{equation} 
with $R_{\textrm{h}}$ being the horizon radius. 
We also have 
that $\textbf{m}>M_{\textrm{\tiny{crit}}}(\textbf{q})$, 
according to the assumptions made in the previous 
subsection. Thanks to the functions $\tilde{N}_0^M(t)$, 
$\tilde{N}_+(t)$,  $\tilde{\Phi}_0(t)$, and $\tilde{\Phi}_+(t)$ 
the Hamiltonian $\textbf{h}$ is an explicitly time 
dependent function. The variational principle associated with the 
reduced action, Eq.\  (\ref{red_action}), will fix the values of 
$\textbf{m}$ and $\textbf{q}$ on the initial and final hypersurfaces, 
or in the spirit of the classical analytical mechanics, 
the Hamiltonian principle fixes 
the initial and final values of the canonical coordinates. 
The equations of motion are 
\begin{eqnarray} \label{red_eom_1} 
  \dot{\textbf{m}} &=& 0\,, \\ 
  \dot{\textbf{q}} &=& 0\,, \label{red_eom_2} \\ 
  \dot{\textbf{p}}_{\textbf{m}} &=& 2\,a(d-2)\tilde{N}_0^M 
  R_\textrm{h}^{d-3} 
  \left\{a(d-2)\left[(d-3)\,k\,R_\textrm{h}^{d-4}+(d-1)l^{-2} 
  R_\textrm{h}^{d-2}-(d-3)\textbf{q}^2R_\textrm{h}^{-(d-2)} 
  \right]\right\}^{-1}\!\!\!\!-\tilde{N}_+\,, 
  \label{red_eom_3}\\ 
  \dot{\textbf{p}}_{\textbf{q}} &=& -(2\,a(d-2))^2\textbf{q}\, 
  \tilde{N}_0^M \left\{a(d-2)\left[(d-3)\,k\,R_\textrm{h}^{d-4}+ 
  (d-1)l^{-2}R_\textrm{h}^{d-2}-(d-3)\textbf{q}^2 
  R_\textrm{h}^{-(d-2)}\right]\right\}^{-1}\!\!+\nonumber \\ 
  &&K_d\,\left(\tilde{\Phi}_0-\tilde{\Phi}_+\right)\,. 
\label{red_eom_4} 
\end{eqnarray} 
The equation of motion for $\textbf{m}$, Eq.\ (\ref{red_eom_1}), is 
understood as saying that $\textbf{m}$ is, on a classical solution, 
equal to a function of the mass parameter of the solution, 
$\textbf{m}=2\,a(d-2)M$, where the solution is given in 
Eqs.\  (\ref{general_metric}) and (\ref{general_f}). 
The same goes for the function $\textbf{q}$, where 
Eq.\  (\ref{red_eom_2}) implies that $\textbf{q}$ is equal to 
the charge parameter on a classical solution, 
Eqs. (\ref{general_metric}) and (\ref{general_f}). 
In order to interpret the other equation of motion, 
Eq.\  (\ref{red_eom_3}), we have to recall that from 
Eq.\  (\ref{p_m}) one has $P_M=-T'$, 
where $T$ is the Killing time. This, together with 
the definition of $\textbf{p}_{\textbf{m}}$, given in 
Eq. (\ref{new_p_m}), yields 
\begin{equation} 
\textbf{p}_{\textbf{m}} = T_0 - T_+\,, 
\end{equation} 
where $T_0$ is the value of the Killing time at the left end of the 
hypersurface of a certain $t$, and $T_+$ is the Killing time at 
spatial infinity, the right end of the same hypersurface of $t$. As 
the hypersurface evolves in the spacetime of the black hole solution, 
the right hand side of Eq.\  (\ref{red_eom_2}) is equal 
to $\dot{T}_0-\dot{T}_+$. Finally, after the definition 
\begin{eqnarray} 
    \textbf{p}_{\textbf{q}} =K_d\,\left(\xi_0 - \xi_+\right)\,, 
\end{eqnarray} 
obtained from Eqs.\ (\ref{vector_potential_plusgauge_tr}), 
(\ref{p_q}), and (\ref{new_p_q}), Eq.\  (\ref{red_eom_4}) 
gives 
$\dot{\textbf{p}}_{\textbf{q}}\propto\dot{\xi}_0 - \dot{\xi}_+$, 
which is the difference of the time derivatives 
of the electromagnetic gauge $\xi(t,r)$ at $r=0$ 
and at infinity. 

\subsection{Quantum theory and partition function} 
\label{q_th_part_function} 

The next step is to quantize the reduced Hamiltonian theory, by 
building the time evolution operator quantum mechanically and then 
obtaining a partition function through the analytic continuation of 
the same operator \cite{louko1}. 
The variables $\textbf{m}$ and $\textbf{q}$ are regarded here 
as configuration variables. These variables satisfy the inequality 
$\textbf{m}>M_{\textrm{\tiny{crit}}}(\textbf{q})$. The 
wave functions will be of the form $\psi(\textbf{m},\textbf{q})$, 
with the inner product given by 
\begin{equation} 
\left(\psi,\chi \right) = \int_A \mu d\textbf{m}d\textbf{q} 
\, \bar{\psi}\chi\,, 
\end{equation} 
where $A$ is the domain of integration defined by 
$\textbf{m}>M_{\textrm{\tiny{crit}}}(\textbf{q})$ and 
$\mu(\textbf{m},\textbf{q})$ is a smooth and positive weight 
factor for the integration measure. It is assumed that $\mu$ 
is a slow varying function, otherwise arbitrary. We are thus 
working in the Hilbert space defined as 
$\mathscr{H}:=L^2(A;\mu d\textbf{m}d\textbf{q})$. 

The Hamiltonian operator, written as $\hat{\textbf{h}}(t)$, acts 
through pointwise multiplication by the function 
$\textbf{h}(\textbf{m},\textbf{q};t)$, 
which on a function of our working Hilbert space reads 
\begin{equation} 
\hat{\textbf{h}}(t)\psi(\textbf{m},\textbf{q})= 
\textbf{h}(\textbf{m},\textbf{q};t) 
\psi(\textbf{m},\textbf{q})\,. 
\end{equation} 
This Hamiltonian operator is an unbounded essentially self-adjoint 
operator. The corresponding time evolution operator in the same 
Hilbert space, which is unitary due to the fact that the Hamiltonian 
operator is self-adjoint, is 
\begin{equation} 
\hat{\textbf{K}}(t_2;t_1) = \exp 
\left[-i\int_{t_1}^{t_2}dt'\,\hat{\textbf{h}}(t') \right]\,. 
\label{K} 
\end{equation} 
This operator acts also by pointwise multiplication in the Hilbert 
space. 
We now define 
\begin{eqnarray} \label{t} 
\mathcal{T} &:=& \int_{t_1}^{t_2}dt\,\tilde{N}_+(t)\,,\\ 
\Theta &:=& \int_{t_1}^{t_2}dt\,\tilde{N}^{M}_0 (t)\,, 
\label{theta}\\ 
\Xi_0 &:=& \int_{t_1}^{t_2}dt\,\tilde{\Phi}_0(t)\,, 
\label{csi_0}\\ 
\Xi_+ &:=& \int_{t_1}^{t_2}dt\,\tilde{\Phi}_+(t)\,. 
\label{csi_+} 
\end{eqnarray} 
Using (\ref{red_Hamiltonian}), (\ref{K}), (\ref{t}), 
(\ref{theta}), (\ref{csi_0}), and (\ref{csi_+}) 
we write the function $K$, which is in fact the 
action of the operator in the Hilbert space, as 
\begin{equation} 
\textbf{K}\left(\textbf{m};\mathcal{T},\Theta,\Xi_0,\Xi_+\right)= 
\exp\left[-i\textbf{m}\,\mathcal{T}+ 2\,i\,a\,R_{\textrm{h}}^{d-2}\, 
\Theta-i\,K_d\,\textbf{q}\,(\Xi_+-\Xi_0) 
\right]\,. 
\label{K2} 
\end{equation} 
This expression indicates that $\hat{\textbf{K}}(t_2;t_1)$ 
depends on $t_1$ and $t_2$ only through the functions 
$\mathcal{T}$, $\Theta$, $\Xi_0$, and $\Xi_+$. Thus, the 
operator corresponding to the function $\textbf{K}$ can now 
be written as $\hat{\textbf{}K}(\mathcal{T},\Theta,\Xi_0,\Xi_+)$. 
The composition law in time 
$\hat{\textbf{K}}(t_3;t_2)\hat{\textbf{K}}(t_2;t_1)= 
\hat{\textbf{K}}(t_3;t_1)$ 
can be regarded as a sum of the parameters 
$\mathcal T$, $\Theta$, $\Xi_0$, and $\Xi_+$ 
inside the operator $\hat{\textbf{K}} 
(\mathcal{T},\Theta,\Xi_0,\Xi_+)$. 
These parameters are evolutions parameters defined by the 
boundary conditions, i.e., $\mathcal{T}$ is the Killing time 
elapsed at right spatial infinity and $\Theta$ is the boost 
parameter elapsed at the bifurcation circle; $\Xi_0$ and $\Xi_+$ 
are line integrals along timelike curves of constant $r$, and 
constant angular variables, at $r=0$ and at infinity. 

\section{Thermodynamics} 
\label{thermo} 

\subsection{Generalities} 
\label{generalities} 
We can now build the partition function for this system. 
The path to follow is to continue the operator to imaginary 
time and take the trace over a complete orthogonal basis. 
Our classical thermodynamic situation consists of a $d$-dimensional 
spherical, planar, or hyperbolic 
charged black hole, asymptotically AdS, 
in thermal equilibrium with a bath of Hawking radiation. Ignoring 
back reaction from the radiation, the geometry is described by the 
solutions in Eqs.\ (\ref{general_metric})-(\ref{vector_potentialform}) 
and (\ref{general_f})-(\ref{vector_potential}). 
Thus, we consider a thermodynamic ensemble in which the temperature, 
or more appropriately here, the inverse temperature 
$\beta$ is fixed, as well as the scalar electric potential $\phi$. 
This characterizes a grand canonical ensemble, 
and the partition function 
$\mathcal{Z}(\beta,\phi)$ arises naturally in such an ensemble. 
To analytically continue the Lorentzian solution 
we put $\mathcal{T}=-i\beta$, and $\Theta=-2\pi i$, this latter choice 
based on the regularity of the classical Euclidean solution. 
We also choose $\Xi_0=0$ and $\Xi_+=i\beta\phi$. 

We arrive then at the following expression for the 
partition function 
\begin{equation} \label{partition_function_1} 
\mathcal{Z}(\beta,\phi) = \textrm{Tr} \left[\hat{K} 
(-i\beta,-2\pi i,0,i\beta\phi)\right]\,. 
\end{equation} 
{}From Eq.\  (\ref{K2}) this is realized as 
\begin{equation} \label{partition_function_2} 
\mathcal{Z}(\beta,\phi) = \int_A \mu\,d\textbf{m}d\textbf{q}\, 
\exp\left[-\beta (\textbf{m}-K_d\,\textbf{q}\phi)+4\,a\,\pi 
R_{\textrm{h}}^{d-2}\right]\left\langle 
\textbf{m}|\textbf{m} 
\right\rangle\,. 
\end{equation} 
Since $\left\langle \textbf{m}|\textbf{m}\right\rangle$ is equal to 
$\delta (0)$, one has to regularize (\ref{partition_function_2}). 
Again, following the Louko-Whiting procedure \cite{louko1}, 
we have to regularize and normalize the operator 
$\hat{\textbf{K}}$ beforehand. This leads to 
\begin{equation} \label{partition_function_3} 
\mathcal{Z}_{\textrm{ren}}(\beta,\phi) = \mathcal{N} \int_{A} 
\mu\,d\textbf{m}d\textbf{q}\,\exp\left[-\beta 
(\textbf{m}-K_d\,\textbf{q}\phi)+ 
4\,a\,\pi R_{\textrm{h}}^{d-2}\right]\,, 
\end{equation} 
where $\mathcal{N}$ is a normalization factor and $A$ is the domain 
of integration.  Provided the weight 
factor $\mu$ is slowly varying compared to the 
exponential in Eq.\  (\ref{partition_function_3}), and using the 
fact that the horizon radius $R_{\textrm{h}}$ is function of 
$\textbf{m}$ and $\textbf{q}$, 
the integral in Eq.\  (\ref{partition_function_3}) is 
convergent. Changing integration variables, from 
$\textbf{m}$ to $R_{\textrm{h}}$, where 
\begin{equation} \label{m_r_h} 
\textbf{m} = a(d-2)R_{\textrm{h}}^{d-3} 
\left(l^{-2}R_{\textrm{h}}^2+k+\textbf{q}^2 
R_{\textrm{h}}^{-2(d-3)}\right)\,, 
\end{equation} 
the integral Eq.\  (\ref{partition_function_3}) becomes 
\begin{equation} \label{partition_function_ren} 
\mathcal{Z}_{\textrm{ren}}(\beta,\phi) = \mathcal{N} \int_{A'} 
\widetilde{\mu}\,dR_{\textrm{h}}d\textbf{q}\,\exp(-I_*)\,, 
\end{equation} 
where $A'$ is the new domain of integration after 
changing variables, and the function 
$I_*(R_{\textrm{h}},\textbf{q})$, a kind of an 
effective action (see \cite{york1}), is written as 
\begin{equation} \label{eff_action} 
I_*(R_{\textrm{h}},\textbf{q}):= \beta\, 
a(d-2)R_{\textrm{h}}^{d-3} 
\left(l^{-2}R_{\textrm{h}}^2+k+\textbf{q}^2 
R_{\textrm{h}}^{-2(d-3)}\right) 
-\beta K_d\,\textbf{q}\,\phi-4\,a\,\pi 
R_{\textrm{h}}^{d-2}\,. 
\end{equation} 
The domain of integration, $A'$, is defined by the inequalities 
$0\leq R_{\textrm{h}}$ and 
$\textbf{q}^2\leq R_{\textrm{h}}^{2(d-3)} 
\left(k+\frac{d-1}{d-3}l^{-2}R_{\textrm{h}}^2\right)$. 
The new weight factor $\widetilde{\mu}$ includes the Jacobian of the 
change of variables. Since the weight factor is slowly varying, 
we can estimate the integral of $\mathcal{Z}_{\textrm{ren}}(\beta,\phi)$ 
by the saddle point approximation. 

\subsection{Effective action and critical points 
(i.e., the solutions of the system)} 
\label{criticalpoints} 

Now we calculate the critical points of the effective action 
(\ref{eff_action}) in order to evaluate the integral of the partition 
function (\ref{partition_function_ren}) through the standard 
saddle-point method.  We need the critical points because the saddle 
point method requires the Taylor expansion of the effective action 
around one of them.  The critical points are found by taking the first 
derivative of the effective action with respect to $R_{\textrm{h}}$ 
and to $\textbf{q}$ and making them zero.  With this, the Taylor 
expansion in the first three terms has only the zeroth order term, the 
effective action evaluated at the critical points, and the second 
order terms evaluated at the critical points as well. Higher orders 
are ignored. This evaluation is done in a close neighborhood of a 
critical point, thus making the zeroth order term more important than 
the second order term. Which one of the critical points is chosen for 
this approximation is seen below. 

The critical points are found through equating to zero 
the first derivatives of the effective action (\ref{eff_action}), 
with respect to $R_{\textrm{h}}$ and to $\textbf{q}$, i.e., 
\begin{equation} 
\label{critpointsdef} 
\frac{\partial I_*}{\partial R_{\textrm{h}}}=0= 
\frac{\partial I_*}{\partial \textbf{q}} 
\end{equation} 
Two distinct sets of critical points are then found: 

(i) The critical points are given as a pair of values 
$(R_{\textrm{h}},\textbf{q})$ 
\begin{eqnarray} 
    R_{\textrm{h}}^\pm &=& \frac{2\pi\,l^2}{(d-1)\beta} 
    \left(1\pm 
    \sqrt{1+\frac{\beta^2(d-3)(d-1)}{16\,a^2\pi^2l^2(d-2)^2} 
    \left\{K_d^2\,\phi^2-\,k\,(2\,a(d-2))^2\right\}}\right) 
    \,, \label{crit_r}\\ 
    \textbf{q}^\pm&=& K_d \frac{\,\phi\, 
    (R_{\textrm{h}}^\pm)^{d-3}}{2\,a(d-2)}\,, 
    \label{crit_q} 
\end{eqnarray} 
where, as a reminder, $K_d=4\,a\sqrt{2\pi\,(d-2)(d-3)}$. 

(ii) Equation (\ref{eff_action}) also has a critical point at 
\begin{eqnarray} 
    R_{\textrm{h}}^{d-4} &=& 0 
    \,, \label{crit_r2}\\ 
    \textbf{q}&=& 0\,, 
    \label{crit_q2} 
\end{eqnarray} 
which for $d>4$ is equivalent to 
$(R_{\textrm{h}},\textbf{q})=(0,0)$. 
For $d=4$ this critical point does not 
exist, but in practice in the study 
of global minima this makes no difference. 

The critical points belong to the domain of 
the effective action (\ref{eff_action}), and so the effective 
action has a derivative equal to zero at these points 
in Eqs.\  (\ref{crit_r})-(\ref{crit_q2}). 
{}From (\ref{crit_r}), if 
\begin{eqnarray} 
    K_d^2\phi^2 &\geq&-\frac{16\,a^2\pi^2l^2(d-2)^2} 
    {\beta^2(d-3)(d-1)} 
    +\,k\,(2\,a(d-2))^2\,, 
    \label{ineq_1} 
\end{eqnarray} 
one finds there is at least one critical point. In more detail 
we find the following: if in Eq.\  (\ref{ineq_1}) there were a $<$ 
instead of a $\geq$, then there would be no critical points. 
If the equality is satisfied in Eq.\  (\ref{ineq_1}), then 
there is only the critical point given by 
$R_{\textrm{h}}=(2\pi\,l^2)/((d-1)\beta)$ and the 
corresponding value of the charge, by replacing 
$R_{\textrm{h}}$ in Eq.\  (\ref{crit_q}). 
In this last case $R_{\textrm{h}}^+=R_{\textrm{h}}^-$. 
If the inequality holds in Eq.\  (\ref{ineq_1}), then 
there are three situations (a) 
$K_d^2\phi^2<k\,(2\,a(d-2))^2$, 
(b) $K_d^2\phi^2=k\,(2\,a(d-2))^2$, and 
(c) $K_d^2\phi^2>k\,(2\,a(d-2))^2$. In (a) there are two 
critical points, given by $(R_{\textrm{h}}^\pm, 
\textbf{q})^\pm$ in (\ref{crit_r})-(\ref{crit_q}), 
in (b) there are two critical points given by 
$R_{\textrm{h}}^+=(4\,\pi\,l^2)/((d-1)\beta)$ and 
$R_{\textrm{h}}^-=0$, with the corresponding 
$\textbf{q}^\pm$ given in (\ref{crit_q}), and 
finally in (c) we have only the critical point 
given by $(R_{\textrm{h}}^+,\textbf{q}^+)$, 
because $R_{\textrm{h}}^-<0$ is not 
physically relevant. All this discussion neglects the critical point 
at $(0,0)$ which is independent of the value of $\phi$, which must be 
added to the discrimination above. One must also note that the 
topology was not taken into account in detail.  If 
the topology is taken into account, one sees 
that for $k=0,-1$ there is always only one critical point of physical 
significance, given by $(R_{\textrm{h}}^+,\textbf{q}^+)$ as $\phi^2$ 
is positive and, for $k=0,-1$, the right hand side of 
Eq.\  (\ref{ineq_1}) is negative.  In this case the critical point at 
$(R_{\textrm{h}}^-,\textbf{q}^-)$ is outside the domain of physical 
interest as $R_{\textrm{h}}^-<0$. 

We can now write the action evaluated at the critical points, where 
also $\beta$ and $\phi$ are determined as functions of the critical 
pair $(R_{\textrm{h}}^{\pm},\textbf{q}^{\pm})$. The full expression is 
\begin{eqnarray} 
I_* = 4\pi a {R_{\textrm{h}}^{\pm}}^{(d-2)} 
\left(\frac{k\,{R_{\textrm{h}}^{\pm}}^{2(d-3)}-{\textbf{q}^{\pm}}^2 
-l^{-2}{R_{\textrm{h}}^{\pm}}^{2(d-2)}} 
{(d-3)\,k\,{R_{\textrm{h}}^{\pm}}^{2(d-3)}- 
(d-3){\textbf{q}^{\pm}}^2+ 
(d-1)l^{-2}{R_{\textrm{h}}^{\pm}}^{2(d-2)}} \right)\,, 
\label{eff_action_at_critpts} 
\end{eqnarray} 
where $a={\Sigma^k_{d-2}}/{16\pi}$, see Eq.\ (\ref{areal}). 
Expression (\ref{eff_action_at_critpts}) is quite general. 
It can be compared with other studies that have chosen a 
particular dimension of spacetime, or a particular asymptotic regime, 
or a particular topology. 

For instance, for generic $d$ and $k=1$ one finds from 
(\ref{eff_action_at_critpts}) an action for $d$-dimensional 
Reissner-Nordstr\'om black holes which was not explicitly shown in 
\cite{cejm1,cejm2}.  If, for generic $d$ and $k=1$, we further put 
$\textbf{q}=0$ one finds from (\ref{eff_action_at_critpts}) the 
following action $I_*=4\pi a {R_{\textrm{h}}^{\pm}}^{(d-2)} 
\left({R_{\textrm{h}}^{\pm}}^{2(d-3)}- 
l^{-2}{R_{\textrm{h}}^{\pm}}^{2(d-2)}\right) 
/\left((d-3)\,{R_{\textrm{h}}^{\pm}}^{2(d-3)}+ 
(d-1)l^{-2}{R_{\textrm{h}}^{\pm}}^{2(d-2)}\right)$ which is the action 
found in \cite{witten}.  In addition, by putting $d=4$ and $k=1$ into 
(\ref{eff_action_at_critpts}) one finds $ I_* =\pi 
R_{\textrm{h}}^2\left( R_{\textrm{h}}^2- \textbf{q}^2- 
l^{-2}R_{\textrm{h}}^4 \right)/\left(R_{\textrm{h}}^2-\textbf{q}^2 
+3l^{-2} R_{\textrm{h}}^4\right)$ which is the action first found in 
\cite{peca1}.  If we further assume $\textbf{q}=0$, one finds the 
Hawking-Page action $I_* = {\pi R_{\textrm{h}}^2\left( 
R_{\textrm{h}}^2- l^{-2}R_{\textrm{h}}^4 
\right)}/\left(R_{\textrm{h}}^2+3l^{-2} R_{\textrm{h}}^4\right)$ 
\cite{hawkingpage}.  Or, by putting $d=4$ and $k=0$ into 
(\ref{eff_action_at_critpts}) one finds $ I_* = \pi 
R_{\textrm{h}}^2\left( \textbf{q}^2+ l^{-2}R_{\textrm{h}}^4 
\right)/\left(\textbf{q}^2-3l^{-2} R_{\textrm{h}}^4\right)$, which was 
found in \cite{peca2}. We defer a full comparison to the next subsection,
after having studied the phase transitions of the system.

\subsection{The most stable solutions and phase transitions} 
\label{stablesolutionsandphasetransitions} 

\subsubsection{Analysis in $d$ dimensions} 
Now, the ensemble in this case has already been identified as the 
grand canonical, where $\textbf{T}$ (or $\beta$) and $\phi$ are 
fixed. What now is needed is to obtain the most stable thermodynamic 
solutions for this ensemble.  In order to do that, it is necessary to 
know the interval of the order parameter $\phi$ in which a given 
critical point is a global minimum, local minimum or none of the 
latter. Each different minimum represents a physical system, such as 
hot flat space, for instance.  Note that, hot flat spaces and 
black holes, are two different sectors of the solution space, one with 
trivial topology, the other with black hole topology, respectively.  To 
be definitive, let us fix the temperature, or $\beta$, on the boundary. 
Now, $\phi$ is also fixed on the boundary, but we can fix it with any 
value we like.  So, imagining changing the parameter $\phi$, fixed on 
the boundary, one finds that, for instance, one local minimum may turn 
into a global minimum, signaling that a phase transition occurs at a 
certain value of $\phi$. 

The saddle point approximation of the 
partition function (\ref{partition_function_ren}) chooses the global 
minima of the critical points for each value of $\phi$, on 
the understanding that $\beta$ is fixed. 
{}From Eqs.\ (\ref{crit_r}), (\ref{crit_q}), (\ref{crit_r2}), and 
(\ref{crit_q2}) one finds the values of the critical 
points in terms of $R_{\textrm{h}}$ and $\textbf{q}$. 
Eqs.\ (\ref{crit_r2}) and (\ref{crit_q2}) are at the origin of the 
$(R_{\textrm{h}},\textbf{q})$, and do not depend on the parameter 
$\phi$. Eqs.\ (\ref{crit_r}) and (\ref{crit_q}) 
will change in value as $\phi$ changes. 
The expression (\ref{crit_q}) implies that once $R_{\textrm{h}}$ 
is determined for a critical point, $\textbf{q}$ is immediately given. 
So, on the plane $(R_{\textrm{h}},\textbf{q})$, 
the critical points are found by the curve defined through 
(\ref{crit_q}), for each given $\phi$. 
Then, if we replace the value of the charge as a function of 
$R_{\textrm{h}}$, i. e., $\textbf{q}(R_{\textrm{h}})$ 
as given in (\ref{crit_q}), into the effective action (\ref{eff_action}) 
we obtain a one variable function of $R_{\textrm{h}}$, 
written as 
\begin{eqnarray} 
    I_*(R_{\textrm{h}})&=&\beta\,a(d-2) 
    \left(l^{-2}R_\textrm{h}^{d-1}+ 
    k\,R_\textrm{h}^{d-3}+K_d^2\phi^2R_\textrm{h}^{d-3} 
    (2\,a(d-2))^{-2}\right)-\nonumber \\ 
    && \beta(2\,a(d-2))^{-1}K_d^2\phi^2R_\textrm{h}^{d-3}- 
    4\,\pi\,aR_\textrm{h}^{d-2} 
    \,, \label{eff_action_r} 
\end{eqnarray} 
which is a function of $R_{\textrm{h}}$. 
Now, not all $R_{\textrm{h}}$ are solutions, i.e., critical points, 
only those defined by (\ref{crit_r}) belong to the 
solution manifold. Choosing from those that belong to 
the solution manifold we are interested in solutions that 
minimize the effective action. 
In order to find out which is a minimum, for a given $\phi$, and even 
which is the lowest minimum of them all, it suffices to analyze the 
one variable function given in (\ref{eff_action_r}). This function 
could have been written with respect to $\textbf{q}$, thus allowing us 
to find the minima in $\textbf{q}$. However, as we are dealing with a 
black hole whose entropy we also want to determine, and as the entropy 
is proportional to the horizon area, which is itself a function of 
$R_{\textrm{h}}$, it is much more convenient to study the effective 
action as a function of $R_{\textrm{h}}$.  Now, the first critical 
point is at $R_{\textrm{h}}=0$.  At this point the action 
(\ref{eff_action_r}) is equal to zero, $I_*(0)=0$.  This critical 
point is always there, as $R_{\textrm{h}}=0$ does not depend on any 
one parameter, except perhaps the dimension, which is considered here 
as $d\geq4$. If $d=4$ then $R_{\textrm{h}}=0$ is no longer a critical 
point, but for any other critical point to be a global minimum it will 
have to obey that $I_*(R^+_{\textrm{h}})<0$. This is valid for all 
the other dimensions where $R_{\textrm{h}}=0$ is in fact a critical 
point. If $I_*(R^+_{\textrm{h}})>0$, the critical point 
in question is not a global 
minimum. All this reasoning is valid 
when there are no two critical points 
such that $I_*(R^+_{\textrm{h}})<0$. In the case of two critical 
points with $I_*(R^+_{\textrm{h}})<0$, the one that gives 
the least value of 
(\ref{eff_action_r}) is the global minimum. 
So the first step is finding the zeros of (\ref{eff_action_r}). Not 
only do the zeros help finding the global minimum, but also they give 
the intervals for the order parameter $\phi$ for each different global 
minimum. In other words, the zeros help discriminate the physical phases. 
The zeros of the effective action (\ref{eff_action_r}) are given by 
\begin{eqnarray} 
 \label{zero_eff_action_r_1} 
 (R_{\textrm{h}}^0)^{d-3} &=& 0\,,\\ 
 {R_{\textrm{h}}^0}^\pm &=& \frac{2\pi\,l^2}{(d-2)\beta} 
 \left(1\pm 
 \sqrt{1+\frac{\beta^2}{16\,a^2\pi^2l^2}\left\{ 
 K_d^2\phi^2-k\,(2\,a(d-2))^2\right\}} 
 \right) 
 \,. 
 \label{zeros_eff_action_r_2} 
\end{eqnarray} 
The first zero, $R_{\textrm{h}}^0=0$ is independent of $\phi$.  In 
relation to the second zero, there follow different conclusions, 
depending on the value of $\phi^2$ for ${R_{\textrm{h}}^0}^\pm$. 
Indeed, ${R_{\textrm{h}}^0}^\pm$  must respect 
\begin{eqnarray} 
    K_d^2\phi^2&\geq&-16\,a^2\pi^2l^2\beta^{-2}+ 
    k\,(2\,a(d-2))^2\,, 
    \label{ineq_2} 
\end{eqnarray} 
so that at least a zero of the form of 
(\ref{zeros_eff_action_r_2}) exists. 
Then, if $K_d^2\phi^2$ were smaller than the right hand 
side of (\ref{ineq_2}) there would be no new zeros, 
besides $R_{\textrm{h}}^0$. This would mean physically that there 
would be no black hole, and there would only be a heat bath with 
temperature $\beta^{-1}$ 
(see \cite{hawkingpage} for the particular case of 
Schwarzschild-AdS). 
If the equality of 
(\ref{ineq_2}) holds, then we have a double zero at 
$R_{\textrm{h}}=(2\pi\,l^2)/((d-2)\beta)$. 
This point is also a critical point, i.e., it is 
a solution for the system. If we replace 
the equality in (\ref{ineq_2}) back in (\ref{crit_r}), 
we see that $R_\textrm{h}^+=(2\pi\,l^2)/((d-2)\beta)$. 
Here both $R_\textrm{h}=0$ and 
$R_\textrm{h}=(2\pi\,l^2)/((d-2)\beta)$ are 
global minima, so there is a degeneracy. 
Here, there is an equal probability for 
the system to consist of a heat bath alone 
or a charged 
black hole of radius $R_\textrm{h}=(2\pi\,l^2)/((d-2)\beta)$ 
with thermal radiation. 
The equality in (\ref{ineq_2}) marks the threshold from 
which the critical point $R_\textrm{h}^+$ is a global 
minimum. Once the inequality in (\ref{ineq_2}) holds, 
then $R_\textrm{h}^+$ is the only global minimum, 
ending the degeneracy, i.e., 
$I_*(R_\textrm{h}^+)<I_*(0)=0$. 
This implies the stable solution 
consists of a charged black hole of radius $R_\textrm{h}^+$ in 
equilibrium with a heat bath temperature $\beta^{-1}$. 
Perhaps some other minor comments in relation to 
$R_\textrm{h}^-$ are in order. 
The other critical point $R_\textrm{h}^-$ is always less than 
$R_\textrm{h}^+$, and the effective action (\ref{eff_action_r}) is 
always positive at $R_\textrm{h}^-$ as long as 
$K_d^2\phi^2-k\,(2\,a(d-2))^2>0$. Once $K_d^2\phi^2-k\,(2\,a(d-2))^2=0$, 
then both the critical point $R_\textrm{h}^-=0$ and 
$I_*(R_\textrm{h}^-)=0$. 
At the same time ${R_{\textrm{h}}^0}^-=0$. However, the action 
(\ref{eff_action_r}) at $R_\textrm{h}^-$ is always larger than at 
$R_\textrm{h}^-$, once there are two different critical points 
$R_\textrm{h}^\pm$. If any critical point $R_\textrm{h}^\pm$ is 
less than zero, it has lost its physical meaningfulness. 

All the previous discussion is general, and applies directly to the 
spherical topology. However, if the topology is different, i.e., 
$k=0,-1$, then the inequality in the expression (\ref{ineq_2}) holds 
always. This means that $R_\textrm{h}^+$ is always the global minimum. 
Therefore the classical solution is found in the global minimum of the 
critical point $(R_{\textrm{h}}^+,\textbf{q}^+)$ of the effective 
action $I_*(R_{\textrm{h}},\textbf{q})$ (Eq.\  (\ref{eff_action})). 
Differently from the spherical case, there are no phase transitions, 
as discussed in \cite{peca2}. 

\begin{figure} [htmb] 
\begin{center} 
\includegraphics[width=6cm,height=3.3cm]{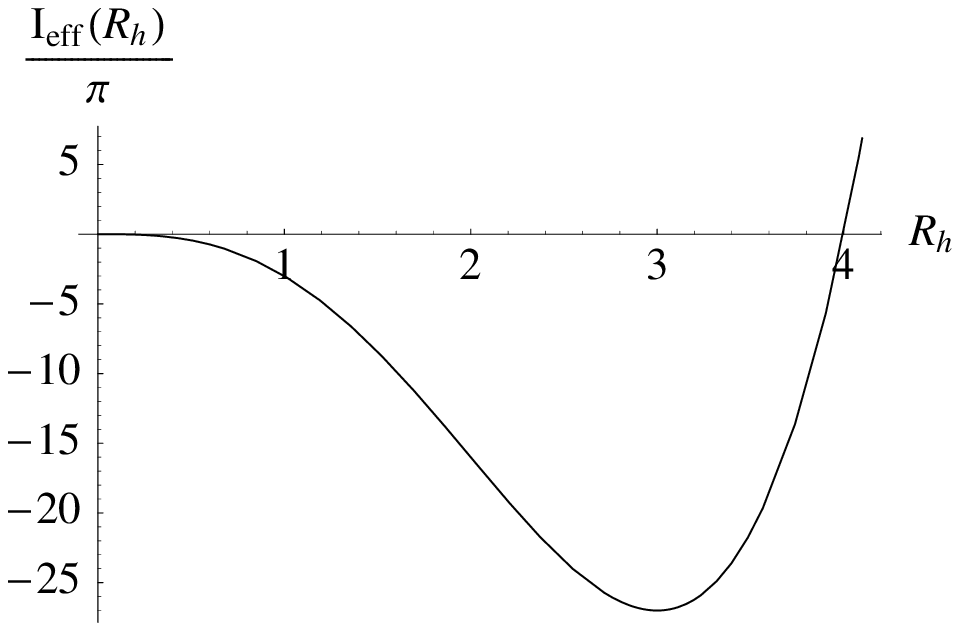} \quad 
\includegraphics[width=6cm,height=3.3cm]{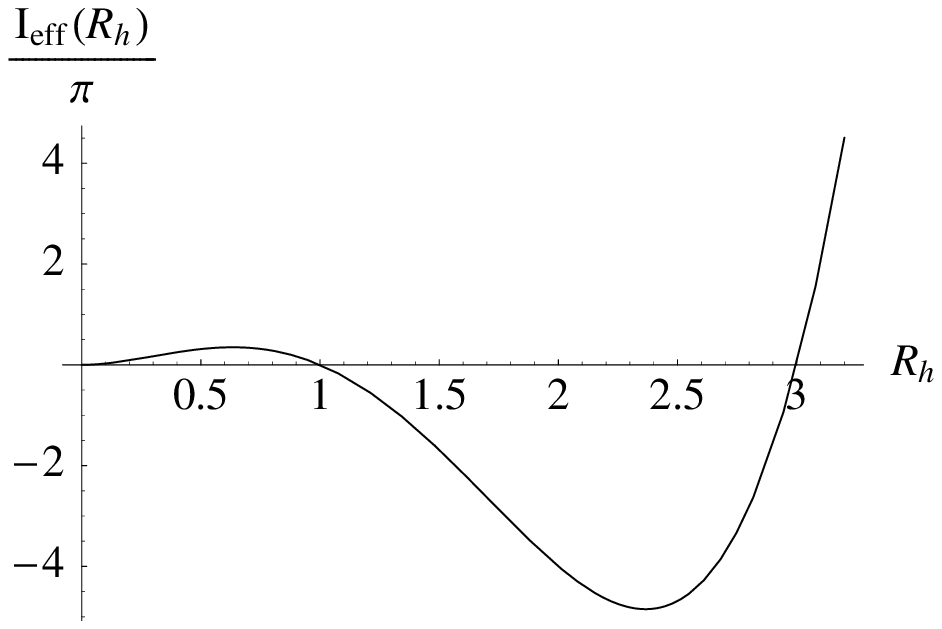} \\ 
($c=0$) 
\qquad\qquad\qquad\qquad 
\qquad\qquad\qquad\qquad\qquad 
($c=3$) 
\end{center} 

\begin{center} 
\includegraphics[width=6cm,height=3.3cm]{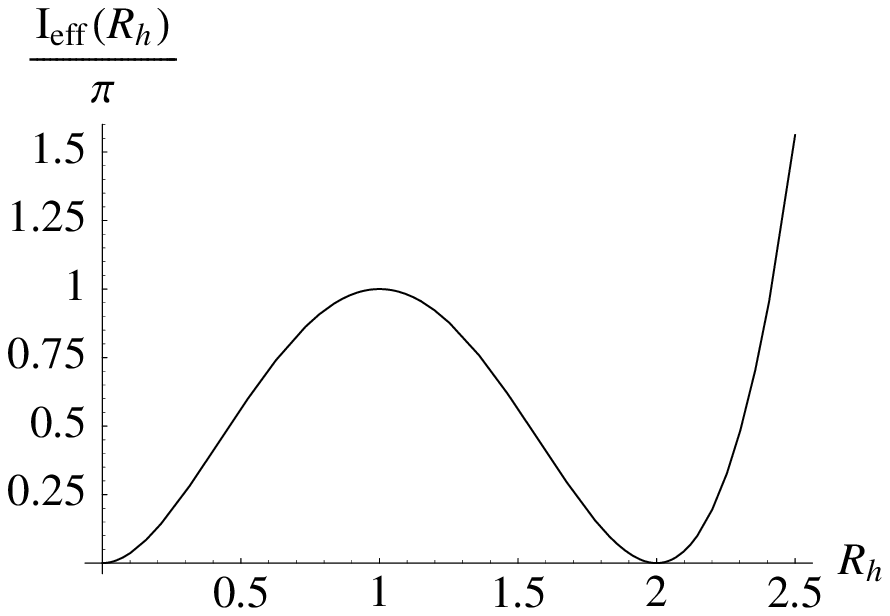} \quad 
\includegraphics[width=6cm,height=3.3cm]{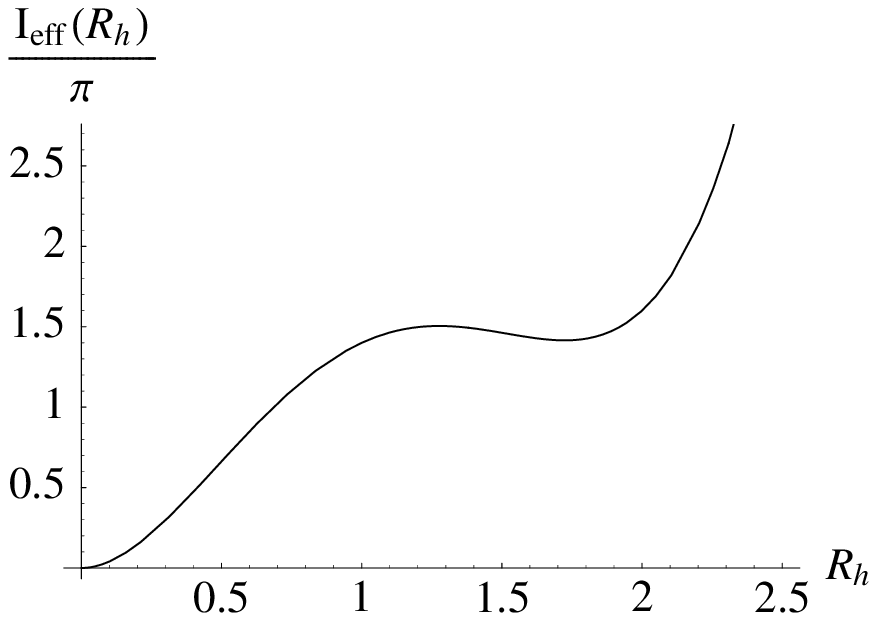}\\ 
($c=4$) 
\qquad\qquad\qquad\qquad 
\qquad\qquad\qquad\qquad\qquad 
($c=4.4$) 
\end{center} 

\begin{center} 
\includegraphics[width=6cm,height=3.3cm]{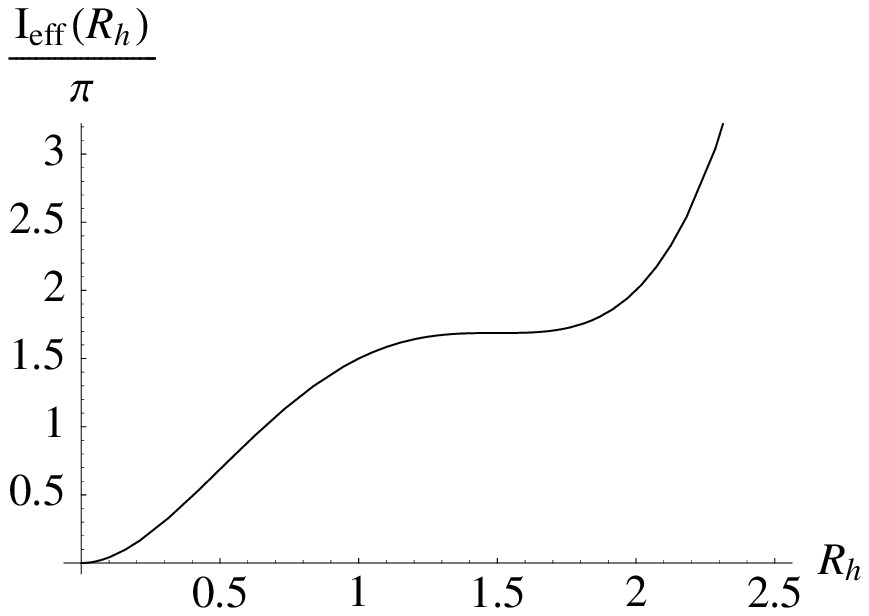} \quad 
\includegraphics[width=6cm,height=3.3cm]{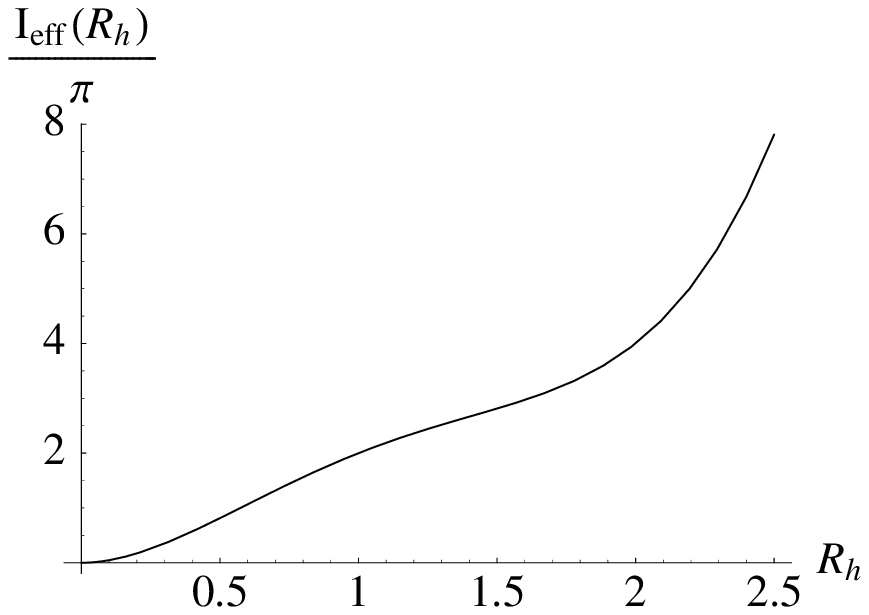}\\ 
(c=$4.5$) 
\qquad\qquad\qquad\qquad 
\qquad\qquad\qquad\qquad\qquad 
($c=5$) 
\end{center} 
\caption{Plots of the effective action, 
$\pi^{-1}I_*(R_\textrm{h})$ in five dimensions, for the choices 
$d=5$, $k=1$, $a=1$, $l=\sqrt{10}$, and 
$c=0,\,3,\,4,\,4.4,\,4.5,\,5$, respectively, 
where $c=10(1-\frac{16}{3}\pi\phi^2)$, see text for details.} 
\label{a} 
\end{figure}

For an illustration of the previous discussion, a choice is made, 
namely, $d=5$, $k=1$, $a=1$, and $l=\sqrt{10}$. 
We also choose $\beta=\frac{10}{3}\pi$. 
Then, the effective action is written in the form 
\begin{eqnarray} 
\pi^{-1}I_*(R_\textrm{h}) &=& 
R_\textrm{h}^4-4R_\textrm{h}^3+cR_\textrm{h}^2\,, 
\label{graphic_eff_action_phi} 
\end{eqnarray} 
where $c=10(1-\frac{16}{3}\pi\phi^2)$. 
So, for a choice of $c=0,\,3,\,4,\,4.4,\,4.5,5$ respectively, 
which amounts to a choice of different values of $\phi$, 
the order parameter for this phase transition, 
we plot, in Figure \ref{a}, 
$\pi^{-1}I_*(R_\textrm{h})$ as a function 
of $R_\textrm{h}$. 
These plots illustrate the process of phase transition from a phase 
where the black hole is a stable solution to a phase where the action 
has a global minimum for hot flat space.  In this instance the phase 
transition happens due to a change in the order parameter $\phi$, the 
electric potential.  The evolution in the plots is toward a higher 
value of $c$, or smaller value of $\phi$.  In more detail, in the 
first plot there is a critical point at a finite radius where the 
effective action is negative. The other critical points are at zero 
radius. Thus the global minimum, the stable solution, is a black hole 
of radius given by the critical point. This is expected on physical 
grounds, since a higher $\phi$ means a higher electrical pressure of 
the walls at infinity on the charged particles which then tend to 
concentrate at the center, forming a black hole.  The next plot shows 
two distinct nonzero critical points, plus the critical point at the 
origin.  The one with larger $R_\textrm{h}$ is the stable solution, 
where again the effective action is negative. In the other critical 
points, the action is either zero or positive. The next plot shows us 
the phase transition. Here the value of $c$ is such that the effective 
action at the larger critical point is equal to zero, as it is zero at 
zero radius. This implies that the effective action has two global 
minima. Here the black hole with the larger radius and hot flat space 
are equally probable. The system is then a mixture of two states which 
can transit from one to the other. The value of $\phi$ is 
$\phi=\frac{3}{\sqrt{80\pi}}$.  For higher values of $c$, the 
likeliest outcome is now hot flat space, because the effective action 
is positive for any value of $c$, except at the origin.  Even so, the 
larger radius critical point is still a local minimum, which allows 
for a metastable black hole solution.  However, there is a limit for 
the existence of such metastable solutions. The last but one plot, for 
$c=4.5$, shows the merging of the two nonzero critical points into 
one, where the effective action is positive and where it has neither a 
maximum nor a minimum. It is thus a completely unstable solution.  For 
values of $c$ such that $c>4.5$ there are no other critical points, 
except for the zero radius, which is equivalent to saying that hot 
flat space is the only outcome of the ensemble. There is thus a value 
of $\phi$ below which there are no black holes for the ensemble. In 
our example the value is 
$\phi=\left(\frac{33}{320\pi}\right)^\frac12$.  This is analogous to 
the Bose-Einstein condensation, which is a phase transition. In this 
case one has as variables the chemical potential $\mu$ and the 
temperature $\textbf{T}$, or $\beta$. If one fixes $\textbf{T}$ and 
increases the number of particles and so increases $\mu$ one has a 
phase transition to a condensate. So $\mu$ is equivalent to $\phi$ 
here, the difference is that bosons feel an effective attractive 
force, whereas charged particles with the same sign feel a repulsive 
force. So, when one lowers $\phi$ one has hot flat space, whereas when 
one raises $\mu$ one has a condensate.  When one raises $\phi$ one has 
a black hole, whereas when one lowers $\mu$ one has a free gas. 

\begin{figure} [htmb] 
\begin{center} 
\includegraphics[width=6cm,height=3.3cm]{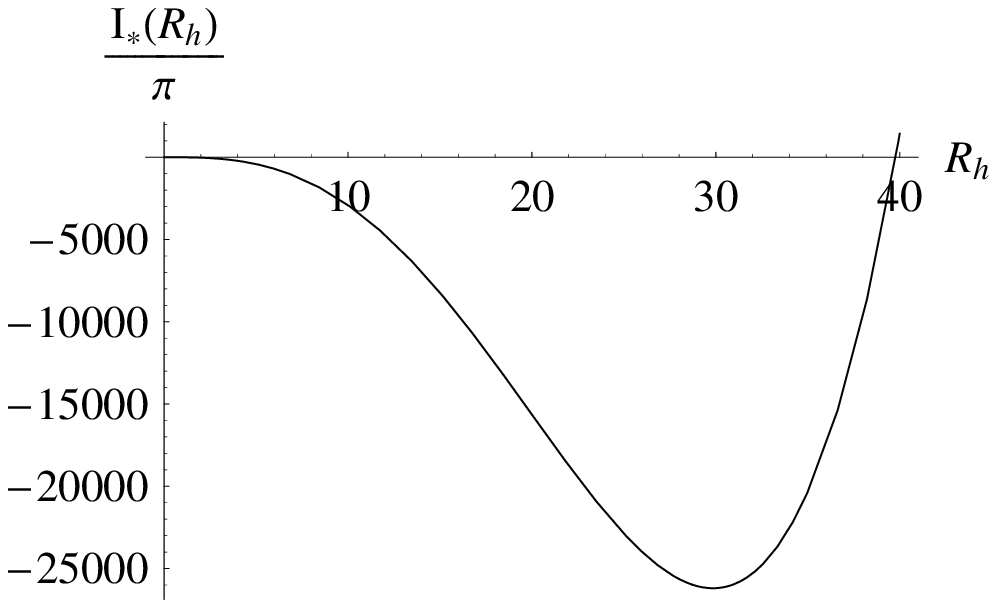} \quad 
\includegraphics[width=6cm,height=3.3cm]{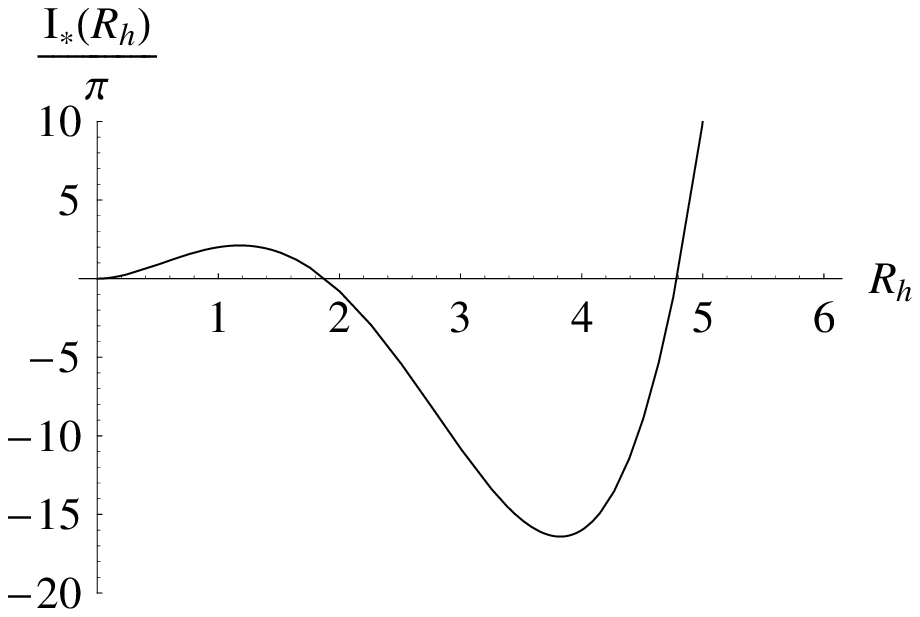} \\ 
($\frac{3\beta}{10\pi}=\frac{1}{10}$) 
\qquad\qquad\qquad\qquad 
\qquad\qquad\qquad\qquad 
($\frac{3\beta}{10\pi}=\frac{6}{10}$) 
\end{center} 

\begin{center} 
\includegraphics[width=6cm,height=3.3cm]{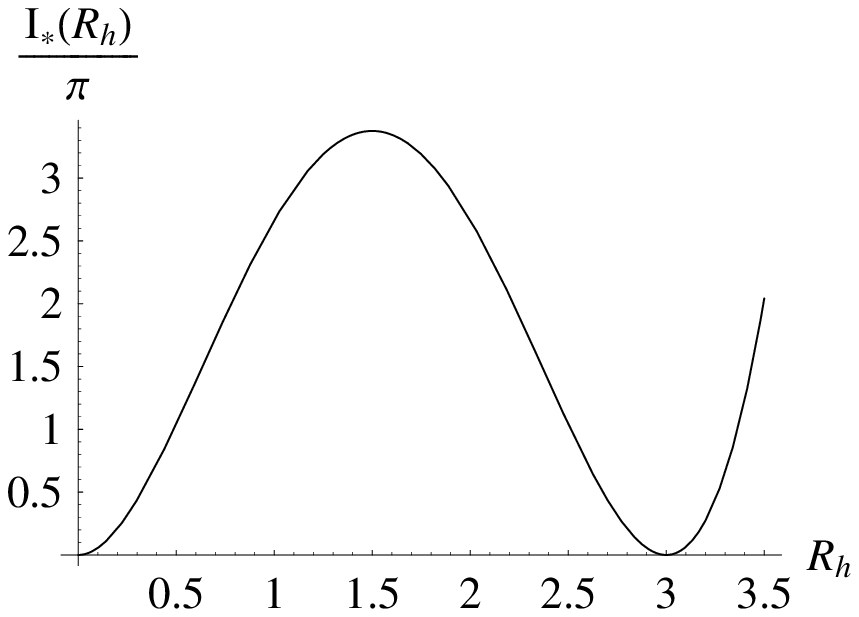} \quad 
\includegraphics[width=6cm,height=3.3cm]{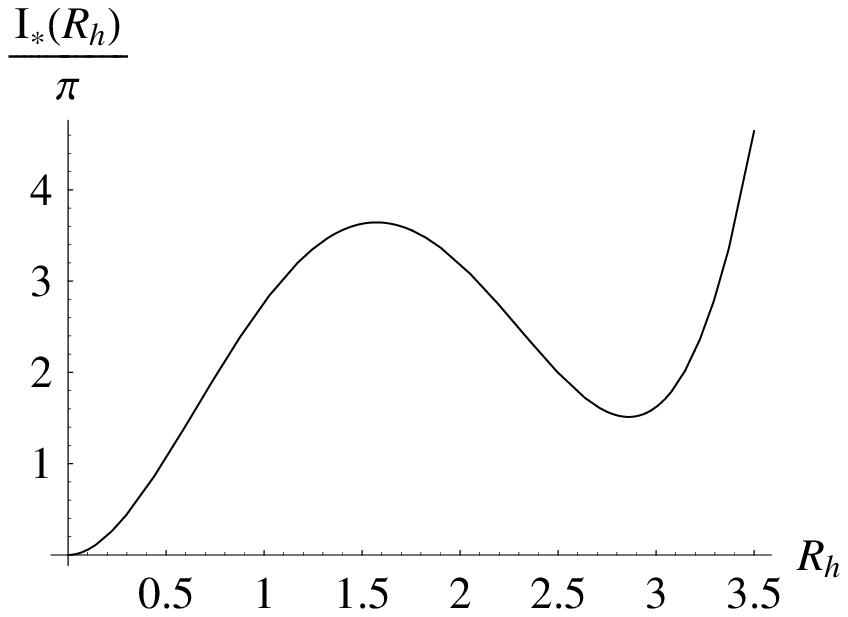}\\ 
($\frac{3\beta}{10\pi}=\frac23$) 
\qquad\qquad\qquad\qquad 
\qquad\qquad\qquad\qquad 
($\frac{3\beta}{10\pi}=\frac23+0.01$) 
\end{center} 

\begin{center} 
\includegraphics[width=6cm,height=3.3cm]{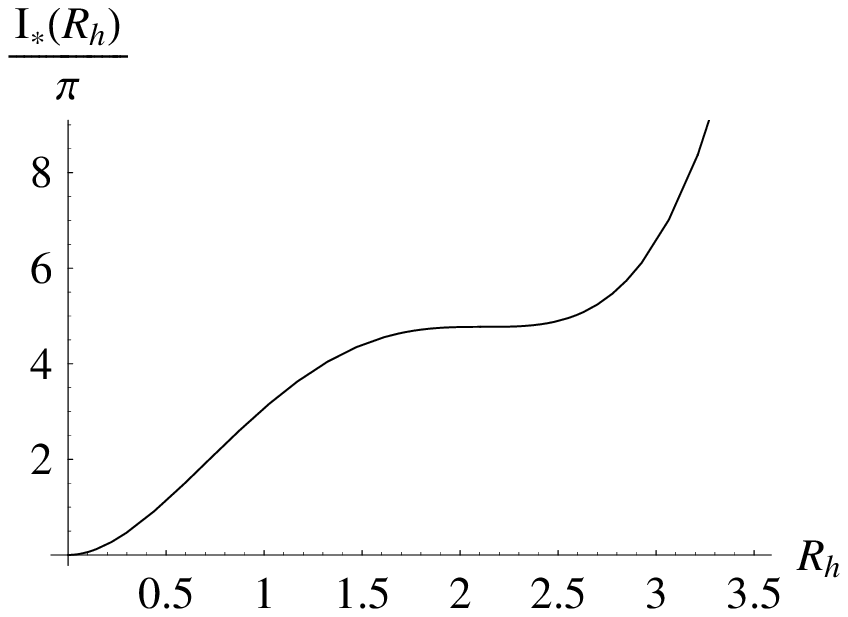} \quad 
\includegraphics[width=6cm,height=3.3cm]{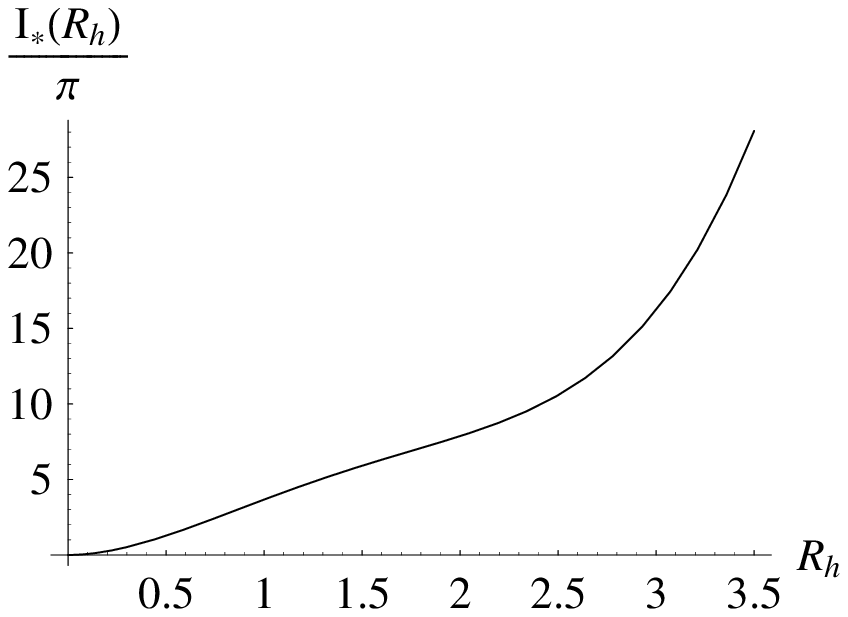} \\ 
($\frac{3\beta}{10\pi}=\frac{\sqrt{2}}{2}$) 
\qquad\qquad\qquad\qquad 
\qquad\qquad\qquad\qquad 
($\frac{3\beta}{10\pi}=\frac23+0.1$) 
\end{center} 
\caption{Plots of the effective action, 
$\pi^{-1}I_*(R_\textrm{h})$ in five dimensions, for the choices 
$d=5$, $k=1$, $a=1$, $l=\sqrt{10}$, and 
$\frac{3\beta}{10\pi}=\frac{1}{10},\,\frac{6}{10},\,\frac23,\,\frac23+0.01, 
\,\frac{\sqrt{2}}{2},\,\frac23+0.1$, respectively, see text for details.} 
\label{b} 
\end{figure}

Now, we can instead study the phase transition by fixing $\phi$ a 
priori, and keeping $\textbf{T}$ , or more precisely, the inverse 
temperature $\beta$, fixed, but allowing it to change from situation 
to situation.  For the same $d=5$, $k=1$, $a=1$, and $l=\sqrt{10}$, 
but now with the value of $\phi$ given a priori by 
$\phi=\frac14\sqrt{\frac{3}{10\pi}}$, the effective actions reduces to 
\begin{eqnarray} 
\pi^{-1}I_*(R_\textrm{h}) &=& \frac{3\beta}{10\pi}R_\textrm{h}^4- 
4R_\textrm{h}^3+\frac{27\beta}{10\pi}R_\textrm{h}^2\,. 
\label{graphic_eff_action_beta} 
\end{eqnarray} 
Through the choice of different values of $\frac{3\beta}{10\pi}$ we 
can make a series of plots which show the different phases of the 
ensemble with respect to the inverse temperature $\beta$. The choices 
for $\frac{3\beta}{10\pi}$ are 
$\frac{3\beta}{10\pi}=\frac{1}{10},\,\frac{6}{10},\,\frac23,\,\frac23+0.01, 
\,\frac{\sqrt{2}}{2},\,\frac23+0.1$.  We plot, in Figure \ref{b}, the 
effective action $\pi^{-1}I_*(R_\textrm{h})$ as a function of 
$R_\textrm{h}$, for those values of $\beta$. 
These plots illustrate the process of phase transition from a phase 
where the black hole is a stable solution to a phase where the action 
has a global minimum for hot flat space.  In this instance the phase 
transition happens due to a change in the order parameter $\beta$, the 
inverse temperature.  The evolution in the plots is toward a higher 
value of $\beta$, or smaller value of the temperature.  In more 
detail, in the first plot there is a critical point at a finite radius 
where the effective action is negative. The other critical points are 
at zero radius. Thus the global minimum, the stable solution, is a 
black hole of a radius given by the critical point.  Note that the 
choice of $\beta=0$ would have yielded an effective action given by 
$I_*(R_\textrm{h})=-4\pi\,R_\textrm{h}^3$, which would not have given 
any critical point except $R_\textrm{h}=0$. This infinite temperature 
ensemble would have had a global maximum at zero radius for the 
effective action, implying that hot flat space at infinite temperature 
is unstable, and at the same time showing that only a black hole of 
infinite radius would satisfy the conditions for stability.  The next 
plot shows two distinct nonzero critical points, plus the critical 
point at the origin.  The larger is the stable solution, where again 
the effective action is negative. In the other critical points, the 
action is either zero or positive. The next plot shows us the 
transition. Here the value of $\beta$ is such that the effective 
action at the larger critical point is equal to zero, as it is zero at 
zero radius. This implies that the effective action has two global 
minima. Here the black hole with the larger radius and hot flat space 
are equally probable. The system is then a mixture of two states which 
can transit from one to the other. The value of $\beta$ is 
$\beta=\frac{20\pi}{9}$.  For higher values of $\beta$, the likeliest 
outcome is now hot flat space, because the effective action is 
positive for any value of $\beta$, except at the origin.  Even so, the 
larger radius critical point is still a local minimum, which allows 
for a metastable black hole solution.  However, there is a limit for 
the existence of such metastable solutions. The last but one plot, for 
$\beta=\frac{10\pi\sqrt2}{6}$, shows the merging of the two nonzero 
critical points into one, where the effective action is positive and 
where it has neither a maximum nor a minimum. It is thus a completely 
unstable solution.  For values of $\beta$ such that 
$\beta>\frac{10\pi\sqrt2}{6}$ there are no other critical points, 
except for the zero radius, which is equivalent to saying that hot 
flat space is the only outcome of the ensemble. 
Again, we can compare this phase transition with the Bose-Einstein 
condensation. Now, one fixes $\phi$ in one case and $\mu$ in the 
other. So when one raises $\beta$ (lowers the temperature) one has hot 
flat space, and a condensate in the other case. When one lowers 
$\beta$ (raises the temperature) one has a black hole, and a free gas 
in the other case. 

We note that there is nothing special happening when we take the 
number of dimensions to be very large, $d\rightarrow\infty$.  All the 
quantities suffer a change in scale, certainly, but qualitatively the 
behavior is analogous. 

\subsubsection{Comparison with other results} 
We can compare our results for $d$-dimensional, Reissner-Nordstr\"om-AdS 
black holes, with either spherical, planar, or hyperbolic horizon 
topology, with results from other authors in several other instances. 
We divide this comparison into the works that also used Hamiltonian 
methods and the works that used Euclidean path integral methods. As we 
have seen, Hamiltonian methods perform a Hamiltonian reduction in the 
Lorentz theory, and then take the trace of an analytically continued 
evolution operator.  Euclidean path integral methods perform a 
Hamiltonian reduction in the analytically continued action. The 
boundary conditions in the two approaches are identical, and the 
difference between them is in the order of quantization (first in the 
Hamiltonian methods) and Euclideanization (first in path integral 
methods). 

\vskip 0.2cm 
(a) Hamiltonian thermodynamic methods 
\vskip 0.1cm 
Firstly, Reissner-Nordstr\"om-AdS black holes in four dimensions 
($d=4$) with spherically symmetric ($k=1$), were studied by Louko and 
Winters-Hilt in \cite{louko3}, using a Hamiltonian thermodynamics 
formalism.  They analyzed these systems in both the canonical ensemble 
and grand canonical ensemble. Now, if we put $d=4$ and $k=1$ in our 
action action (\ref{eff_action}) and in the critical point 
(\ref{crit_r})-(\ref{crit_q}), it reduces then to the grand canonical 
ensemble effective action and the critical point of \cite{louko3}. 
There is a small difference in the coefficient $K_d$ between our work 
and \cite{louko3}, when $d=4$, due to a different choice in the 
coefficient of the Maxwell term in the action 
(\ref{einsteinmaxwell1}). Here we have followed Myers and Perry 
\cite{mp}, where the Maxwell term is $\frac14 F_{\mu\nu}F^{\mu\nu}$, 
whereas in \cite{louko3} there is no $\frac14$ term. 

Secondly, Reissner-Nordstr\"om-AdS black holes in four dimensions 
($d=4$) with planar and hyperbolic symmetry ($k=0,-1$), were studied 
by Brill, Louko and Peld\'an in \cite{louko4} in both, the canonical 
ensemble and grand canonical ensemble, using implicitly a Hamiltonian 
thermodynamics formalism.  When we put $d=4$ and ($k=0,-1$) in our 
action (\ref{eff_action}) and critical point 
(\ref{crit_r})-(\ref{crit_q}) one recovers the results in 
\cite{louko4} for the grand canonical ensemble. Here, as with the case 
above, there is the caveat of the constant $K_d$, i.e., it will not 
render the same numerical coefficient due to the choice of a different 
constant for the Maxwell term in the action (\ref{einsteinmaxwell1}). 

Thirdly, it is interesting to note that the calculations done, also 
using a Hamiltonian thermodynamics formalism, by Louko, Simon, and 
Winters-Hilt in \cite{louko5} and ours match in the appropriate 
limits.  In \cite{louko5} it was studied a spherical black hole 
$(k=1)$ in a finite box with radius $r_{\rm B}$ in a Gauss-Bonnet 
(Lovelock) theory with parameter $\hat{\lambda}$, without cosmological 
constant, in five dimensions, $d=5$. Here we have studied charged 
black holes in general relativity with cosmological constant in $d$ 
dimensions. Thus, in order to match both results, we do the following: 
in \cite{louko5} we take the Gauss-Bonnet parameter equal to zero 
($\hat{\lambda}=0$) to recover general relativity, and take the box to 
infinity ($r_{\rm B}\to\infty$); here we take $d=5$, the cosmological 
constant and the electrical charge equal to zero ($l\to\infty$ and 
$\textbf{q}=0$), and assume spherical topology ($k=1$).  Then one gets 
as the equation for the critical point of the effective action 
$\beta=2\pi R_{\textrm{h}}$, both in \cite{louko5} (see Eq. (5.3) of 
\cite{louko5}), and here (take the appropriate limit in 
(\ref{crit_r})). 

\vskip 0.4cm 
(b) Euclidean path integral methods 
\vskip 0.1cm 

Firstly, the thermodynamics and phase transitions of 
the spherical Reissner-Nordstr\"om black holes 
in AdS spacetimes 
in $d$-dimensions can  be studied if we put $k=1$ 
in Eq.\ (\ref{eff_action_at_critpts}), yielding 
\begin{eqnarray} 
I_* &=& 4\pi a {R_{\textrm{h}}^{\pm}}^{(d-2)} 
\left(\frac{{R_{\textrm{h}}^{\pm}}^{2(d-3)}- 
{\textbf{q}^{\pm}}^2- 
l^{-2}{R_{\textrm{h}}^{\pm}}^{2(d-2)}} 
{(d-3)\,{R_{\textrm{h}}^{\pm}}^{2(d-3)}- 
(d-3){\textbf{q}^{\pm}}^2 
+(d-1)l^{-2}{R_{\textrm{h}}^{\pm}}^{2(d-2)}}\right)\,, 
\label{actioncejm} 
\end{eqnarray} 
the case studied in \cite{cejm1,cejm2} but in which the action 
(\ref{actioncejm}) was not explicitly displayed.  If we further put 
$\textbf{q}=0$ in Eq.\ (\ref{eff_action_at_critpts}), we obtain the 
action for Schwarzschild black holes in AdS spacetimes in 
$d$-dimensions, i.e., 
\begin{eqnarray} 
I_* &=& 4\pi a {R_{\textrm{h}}^{\pm}}^{(d-2)} 
\left(\frac{{R_{\textrm{h}}^{\pm}}^{2(d-3)}- 
l^{-2}{R_{\textrm{h}}^{\pm}}^{2(d-2)}} 
{(d-3)\,{R_{\textrm{h}}^{\pm}}^{2(d-3)}+ 
(d-1)l^{-2}{R_{\textrm{h}}^{\pm}}^{2(d-2)}}\right)\,. 
\label{actionwitten} 
\end{eqnarray} 
This action was found by Witten \cite{witten} through mixed methods. 
On further choosing $d=4$ one finds the Hawking-Page action 
\cite{hawkingpage}. 

Secondly, Reissner-Nordstr\"om-AdS black holes in four dimensions 
($d=4$) with spherically symmetric ($k=1$), were studied by Pe\c ca 
and Lemos in \cite{peca1} using instead Euclidean path integral 
methods in the grand canonical ensemble, and where the black holes are 
put inside a rigid box of radius $r_{\rm B}$.  As already remarked in 
\cite{peca1} the results match, as they should, those of 
\cite{louko3}, when $r_{\rm B}\rightarrow\infty$.  It is interesting 
to further compare \cite{peca1} with our results here.  In 
\cite{peca1}, when taking the limit $r_{\rm B}\to\infty$, it was found 
there are two sets of solutions, which are obtained through different 
ways: (i) fix $r_+$, which is our $R_{\textrm{h}}$, and the charge of 
the black hole $e$, which is our $\textbf{q}$, or (ii) fix the 
boundary conditions, which means fixing $\beta$ and $\phi$.  This 
second set yields, as $r_{\rm B}\to\infty$, a divergent 
$R_{\textrm{h}}\sim r_{\rm B}$ and a divergent $\textbf{q} \sim r_{\rm 
B}^2$, and also leads to an entropy going to infinity as $r_{\rm 
B}^2$.  In this limit the heat capacity and the thermal energy diverge 
too.  These very interesting solutions, first found for the 
asymptotically flat Schwarzschild black holes, in \cite{york1}, 
represent metastable black holes which can quantum tunnel into hot 
flat space and vice versa. Since we 
have taken from the start that our box is AdS infinity (i.e., $r_{\rm 
B}\to\infty$ at the outset), our boundary conditions do not recover 
these metastable solutions, and we will not discuss them anymore (see 
\cite{peca1} for further analysis of these solutions). The first set, 
which fixes $R_{\textrm{h}}$ and $\textbf{q}$, is of our concern now. 
In \cite{peca1}, the temperature $\textbf{T}$ and electrostatic 
potential $\phi$ tend to zero as $r_{\rm B}^{-1}$, and the energy 
tends to zero as well, $E=M/(1+r_{\rm B}^2/l^2)^{\frac12}$.  Here we 
have fixed precisely the temperature at infinity (following 
\cite{louko3}) so that $E=M$. But the difference between the two 
adjustments is merely one of choosing the zero point.  After these 
remarks we can proceed, and by putting $d=4$, $k=1$ into 
(\ref{eff_action_at_critpts}) one finds 
\begin{eqnarray} 
I_* &=& \frac{\pi R_{\textrm{h}}^2\left( R_{\textrm{h}}^2- 
\textbf{q}^2- 
l^{-2}R_{\textrm{h}}^4 \right)}{R_{\textrm{h}}^2-\textbf{q}^2 
+3l^{-2} 
R_{\textrm{h}}^4}\,, 
\label{actionpecalemos} 
\end{eqnarray} 
which is precisely equal to the one first found in \cite{peca1}.  If 
we further put $\textbf{q}=0$, one finds again the Hawking-Page action 
\cite{hawkingpage} $I_* = {\pi R_{\textrm{h}}^2\left( 
R_{\textrm{h}}^2- l^{-2}R_{\textrm{h}}^4 
\right)}/{R_{\textrm{h}}^2+3l^{-2} R_{\textrm{h}}^4}$, an action which 
has attracted much attention.  All these black hole solutions in AdS 
spacetime are stable.  We can also find the action in the 
asymptotically flat case, i.e., $l\to\infty$. Then, for $d=4$, $k=1$, 
and $l\to\infty$ in Eq.\ (\ref{eff_action_at_critpts}) yields $I_* = 
\pi R_{\textrm{h}}^2$, which were first found by York \cite{york1} for 
the uncharged case, and by \cite{braden} for the charged case.  One 
also finds that these asymptotically flat black holes are unstable in 
the grand canonical ensemble, although the branch of large black holes 
not recovered by our boundary conditions are stable (see again 
\cite{york1,braden}, see also \cite{grossperryyaffe} for the 
discussion of the metastability of these objects). 

Thirdly, Reissner-Nordstr\"om-AdS black holes in four dimensions 
($d=4$) with toroidal planar topology ($k=0$), were studied by Pe\c ca 
and Lemos in \cite{peca2} using Euclidean path integral methods in the 
grand canonical ensemble, and where the black holes are put inside a 
rigid box of radius $r_{\rm B}$.  As already remarked in \cite{peca2} 
the results match, as they should, those of \cite{louko5}, for all 
$r_{\rm B}$, including when $r_{\rm B}\rightarrow\infty$.  It is 
interesting to further compare \cite{peca2} with our results here. 
Again, in \cite{peca2}, when taking the limit $r_{\rm B}\to\infty$, it 
was found there are two sets of solutions, which are obtained through 
different ways: (i) fix $R_{\textrm{h}}$ and the charge of the black 
hole $\textbf{q}$, or (ii) fix the boundary conditions, which means 
fixing $\beta$ and $\phi$.  This second set yields, as $r_{\rm 
B}\to\infty$, a divergent $R_{\textrm{h}}\sim r_{\rm B}$ and a 
divergent $\textbf{q} \sim r_{\rm B}^2$, and also leads to an entropy 
going to infinity as $r_{\rm B}^2$.  In this limit the heat capacity 
and the thermal energy diverge too.  As in the spherical case, these 
solutions, represent metastable black holes which can quantum tunnel 
into hot flat space and vice versa.  Since we have taken from the 
start that our box is AdS infinity (i.e., $r_{\rm B}\to\infty$ at the 
outset), our boundary conditions do not recover these metastable 
solutions, and we will not discuss them anymore (see \cite{peca2} for 
further analysis).  The first set, which fixes $R_{\textrm{h}}$ and 
$\textbf{q}$, is of our concern now.  In \cite{peca2}, the temperature 
$\textbf{T}$ and electrostatic potential $\phi$ tend to zero as 
$r_{\rm B}^{-1}$, and the energy tends to zero as well, $E\sim M/(\pi 
r_{\rm B}/l)$.  Here we have fixed precisely the temperature at 
infinity (following \cite{louko3}) so that $E=M$. But the difference 
between the two adjustments is merely one of choosing the zero point. 
After these remarks we can proceed, and by putting $d=4$, $k=0$ into 
(\ref{eff_action_at_critpts}) we find 
\begin{eqnarray} 
I_* &=& \frac{\pi R_{\textrm{h}}^2\left( \textbf{q}^2+ 
 l^{-2}R_{\textrm{h}}^4 \right)}{\textbf{q}^2-3l^{-2} 
R_{\textrm{h}}^4}\,. 
\label{actionpecalemos2} 
\end{eqnarray} 
which was first found in \cite{peca2}. 

The analysis displayed in (a) and (b) above, and the comparisons made, 
show that the new action, Eq.\  (\ref{eff_action}), for general $d$ and 
arbitrary topologies recovers previous results in the appropriate 
limits. 

\subsection{Thermodynamic quantities for 
the solution containing a black hole} 
\label{thermodynamicquantities} 

The partition function $\mathcal{Z}_{\textrm{ren}}(\beta,\phi)$ 
can now be written, for 
$K_d^2\phi^2>-16\,a^2\pi^2l^2\beta^{-2}+k\,(2\,a(d-2))^2$, 
through the saddle-point approximation 
\begin{eqnarray} \label{z_approx} 
    \mathcal{Z}_{\textrm{ren}}(\beta,\phi)&\approx&\textrm{P} 
    \exp\left[-I_*(R_{\textrm{h}}^+,\textbf{q}^+)\right]\,,\\ 
    \ln\mathcal{Z}_{\textrm{ren}}(\beta,\phi) &\approx& \ln\textrm{P} 
    -I_*(R_{\textrm{h}}^+,\textbf{q}^+)\,. 
    \label{log_z_approx} 
\end{eqnarray} 
where $\textrm{P}$ is a slowly varying prefactor, and 
$I_*(R_{\textrm{h}}^+,\textbf{q}^+)$ is the effective action 
evaluated at the global minimum critical point. 
By ignoring the prefactor's logarithm, which closer to 
$(R_{\textrm{h}}^+,\textbf{q}^+)$ is less relevant, 
we are able to determine the 
value of $\textbf{m}$ at the critical point, where we find that it 
corresponds to the value of the mass of the classical solution of the 
black hole given in Eqs.\ (\ref{general_metric})-(\ref{general_f}). 
Thus, when the critical point dominates the partition function, 
we have that the mean energy $\left\langle E \right\rangle$ 
is given by 
\begin{equation} 
\left\langle E\right\rangle = -\frac{\partial}{\partial\beta}\ln 
\mathcal{Z}_{\textrm{ren}} \approx 
a(d-2){R_\textrm{h}^+}^{d-3} 
\left(l^{-2}{R_\textrm{h}^+}^{2}+k\, 
+K_d^2\,\phi^2(2\,a(d-2))^{-2}\right) 
= \textbf{m}^+\,, 
\label{m_saddle_point} 
\end{equation} 
where $\textbf{m}^+$ is obtained from Eq.\  (\ref{m_r_h}) evaluated at 
$(R_{\textrm{h}}^+,\textbf{q}^+)$. 
The thermal expectation value of the charge is 
\begin{equation} 
    \left\langle Q\right\rangle = \beta^{-1}\frac{\partial} 
        {\partial\phi} 
    \ln \mathcal{Z}_{\textrm{ren}}\approx K_d\,\textbf{q}^+\,, 
    \label{q_plus} 
\end{equation} 
as given in Eq.\  (\ref{crit_q}). 
We also can write the temperature $\textbf{T}\equiv\beta^{-1}$ as 
a function of the critical point $R_{\textrm{h}}^+$ and the 
electric charge $\textbf{q}^+$ 
\begin{eqnarray} 
    \textbf{T} &=& 
    \frac{(d-1)\,l^{-2}{R_\textrm{h}^+}^{2}+ 
    (d-3)\,k-(d-3)\,\left(\textbf{q}^+\right)^2 
        {R_\textrm{h}^+}^{-2(d-3)}} 
    {4\pi{R_{\textrm{h}}^+}}\,, \label{temperature} 
\end{eqnarray} 
with $R_{\textrm{h}}^+$ explicitly written in (\ref{crit_r}), and 
where $\textbf{q}^+$ is given in in Eq.\  (\ref{crit_q}).  If we replace 
the value of the charge of the extreme black hole solution, which is 
given in $\textbf{q}^2=R_{\textrm{h}}^{2(d-3)} 
\left(k+\frac{d-1}{d-3}l^{-2}R_{\textrm{h}}^2\right)$, the maximum the 
charge can attain inside the domain of integration of the partition 
function, into the expression for the temperature, Eq.\ 
(\ref{temperature}), 
then we get a null temperature. 
It can be shown that 
$\partial\textbf{m}^+/\partial\beta<0$, which through the constant 
$\phi$ heat capacity $C_{\phi}=-\beta^2(\partial \left\langle 
E\right\rangle/\partial\beta)$ tells us that the system is 
thermodynamically stable. 

Finally there is the entropy of the black 
hole, $S$, which is given by 
\begin{equation} 
S = 
\left(1-\beta\frac{\partial}{\partial\beta}\right) 
\ln\mathcal{Z}_{\textrm{ren}} 
= \frac14 \Sigma^k_{d-2}{R_\textrm{h}^+}^{d-2} 
\,. 
\label{entropy} 
\end{equation} 
We see that the extreme black 
hole, despite having null temperature, has an entropy proportional to 
its horizon area. 

\section{Conclusions} 
\label{conclusions} 

{}From a Hamiltonian thermodynamics formalism we have found a quite 
general effective action (\ref{eff_action_at_critpts}) valid for 
charged higher dimensional $d\geq4$ AdS (negative cosmological 
constant) spacetimes, with spherical, planar, and hyperbolic 
topologies.  There are black hole as well as hot flat spacetime 
solutions.  It was shown that the phase transitions, already present 
for the Reissner-Nordstr\"om AdS black hole in four dimensions, show 
up again for the generic $d$-dimensional charged case. In these phase 
transitions a black hole turns into hot flat space and vice-versa. 
However, this phase transition does not happen for nonspherical 
topologies. We have also calculated the entropy of these higher 
dimensional charged AdS black holes. This entropy is given by a 
quarter of the horizon area for any topology considered. It is thus 
obtained that the entropy relation with the quarter of the horizon 
area is kept for arbitrary higher dimensions, and that the topology 
does not affect the relation. Since the Hamiltonian formalism yields a 
correct consistent expression for the entropy of a black hole, we have 
proved that the formalism has a wide applicability, both to higher 
dimensions and to nonspherical topologies.  These results may be of 
value for investigations on AdS/CFT physics. 

\begin{acknowledgments} 
GASD thanks the support from Funda\c c\~ao para a Ci\^encia e 
Tecnologia (FCT) - Portugal through the fellowship SFRH/BD/2003. 
This work was partially funded by Funda\c c\~ao para a Ci\^encia e 
Tecnologia (FCT) - Portugal, through project 
PPCDT/FIS/57552/2004. 
\end{acknowledgments}


\newpage 
\begin{thebibliography}{99} 
\bibitem{hawking1} S. W. Hawking, Commun. Math. Phys. \textbf{43}, 199 
  (1975). 
\bibitem{bardeencarterhawking1973}  J. M. Bardeen, B. Carter, and 
  S. W. Hawking, Commun. Math. Phys. \textbf{31}, 161 (1973). 
\bibitem{hartlehawking} J. B. Hartle and S. W. Hawking, Phys. Rev. D 
  \textbf{13}, 2188 (1976). 
\bibitem{hawking2} S. W. Hawking, in \emph{General Relativity: An Einstein 
Centenary Survey}, edited by S. W. Hawking and W. Israel (Cambridge 
University Press, Cambridge, England, 1979). 
\bibitem{york1} J. W. York, Phys. Rev. D \textbf{33}, 2092 (1986). 
\bibitem{braden} H. W. Braden, J. D. Brown, B. F. Whiting, and 
  J. W. York, Phys. Rev. D \textbf{42}, 3376 (1990). 
\bibitem{zaslavskii1} O. B. Zaslavskii, Phys. Lett. A 
                   {\bf 152}, 463 (1991). 
\bibitem{hawkingpage} S. W. Hawking and D. N. Page, Commun. Math. Phys. 
\textbf{87}, 577 (1983). 
\bibitem{peca1} C. S. Pe\c{c}a and J. P. S. Lemos, Phys. Rev. D 
  \textbf{59}, 124007 (1999). 
\bibitem{peca2} C. S. Pe\c{c}a and J. P. S. Lemos, J. Math. Phys. 
\textbf{41}, 4783 (2000). 
\bibitem{witten} E. Witten, Adv. Theor. Math. Phys. \textbf{2}, 505 
(1998). 
\bibitem{cejm1} A. Chamblin, R. Emparan, C. V. Johnson, and R. C. Myers, 
           Phys. Rev. D   \textbf{60}, 064018 (1999). 
\bibitem{cejm2} A. Chamblin, R. Emparan, C. V. Johnson, and R. C. Myers, 
           Phys. Rev. D   \textbf{60}, 104026 (1999). 
\bibitem{grossperryyaffe} D. J. Gross, M. J. Perry, and L. G. Yaffe, 
                      Phys. Rev. D \textbf{25}, 330 (1982). 
\bibitem{dirac} P. A. M. Dirac, 
\emph{Lectures on Quantum Mechanics}, (Yeshiva University, New York, 
1964). 
\bibitem{adm} R. Arnowitt, S. Deser, and C. W. Misner, in 
  \emph{Gravitation: an introduction to current research}, edited by 
  L. Witten (Wiley, New York, 1962), Chap. 7, p. 227 (see also 
  arXiv:gr-qc/0405109). 
\bibitem{rt} T. Regge and C. Teitelboim, Ann. Phys. (NY) \textbf{88}, 
  286 (1974). 
\bibitem{kuchar} K. V. Kucha\v{r}, Phys. Rev. D \textbf{50}, 3961 (1994). 
\bibitem{louko1} J. Louko and B. F. Whiting, Phys. Rev. D \textbf{51}, 
  5583 (1995). 
\bibitem{louko3} J. Louko and S. N. Winters-Hilt, Phys. Rev. D 
  \textbf{54}, 2647 (1996). 
\bibitem{louko4} D. R. Brill, J. Louko, and P. Peld\'an, Phys. Rev. D 
  {\bf 56} 3600 (1997). 
\bibitem{louko5} 
  J. Louko, J. Z. Simon, and S. N. Winters-Hilt, Phys. Rev. D {\bf 55} 
  3525 (1997). 
\bibitem{kunst1} G. Kunstatter, R. Petryk, and S. Shelemy, 
  Phys. Rev. D \textbf{57}, 3537 (1998). 
\bibitem{kunst2} A. J. M. Medved and G. Kunstatter, 
  Phys. Rev. D \textbf{59}, 104005 (1999). 

\bibitem{louko2} S. Bose, J. Louko, L. Parker, and Y. Peleg, 
  Phys. Rev. D \textbf{53}, 5708 (1996). 
\bibitem{dl3} G. A. S. Dias and J. P. S. Lemos, Phys. Rev. D {\bf 
78}, 044010 (2008). 
\bibitem{dl4} G. A. S. Dias and J. P. S. Lemos, Phys. Rev. D {\bf 
78}, 084020 (2008). 
\bibitem{bose} S. Bose, L. Parker, and Y. Peleg, Phys. Rev. D 
  \textbf{56}, 987 (1997). 
\bibitem{tangherlini} F. R. Tangherlini, Nuovo Cim. {\bf 27}, 636 
  (1963). 
\bibitem{mp} R. C. Myers and M. J. Perry, 
Ann. Phys. (N.Y.) \textbf{172}, 304 (1986). 
\bibitem{lemos1} J. P. S. Lemos, Class. Quant. Grav. \textbf{12}, 1081 
  (1995). 
\bibitem{lemos2} J. P. S. Lemos, Phys. Lett. B \textbf{353}, 46 (1995). 
\bibitem{lemos_zanchin} J. P. S. Lemos and V. T. Zanchin, 
                        Phys. Rev. D \textbf{54}, 3840 (1996). 
\bibitem{cai1} R. G. Cai and Y. Z. Zhang, Phys. Rev. D 
   \textbf{54}, 4891 (1996). 
\bibitem{vanzo} L. Vanzo, Phys. Rev. D \textbf{56}, 6475 (1997). 
\bibitem{mann} R. B. Mann, Class. Quant. Grav. \textbf{14}, 2927 
  (1997). 
\bibitem{birmingham} D. Birmingham, Class. Quant. Grav. 
                     \textbf{16}, 1197 (1999). 
\bibitem{cai2} R. G. Cai and K. S. Soh, Phys. Rev. D \textbf{59}, 
044013 (1999). 
\bibitem{santos_dias_lemos} N. L. Santos, O. J. C. Dias, and J. P. S. 
         Lemos, Phys. Rev. D  \textbf{70}, 124033 (2004). 
\bibitem{kerrantidesitter} G. W. Gibbons, H. Lu, D. N. Page, and 
  C. N. Pope, J. Geom. Phys. \textbf{53}, 49 (2005). 
\bibitem{ht} M. Henneaux and C. Teitelboim, 
  Commun. Math. Phys. \textbf{98}, 391 (1985). 

\end{thebibliography}
\end{document}